\pdfoutput=1
\documentclass[useAMS]{mn2e}
\usepackage{graphicx}
\usepackage{amssymb}
\usepackage{amsmath}

\usepackage{fixltx2e}

\newcommand\be{\begin{equation}}
\newcommand\ee{\end{equation}}

\title [RWI \& Long-Term Evolution of Dead Zones]{Rossby Wave Instability and Long-Term Evolution of Dead Zones in Protoplanetary Discs}

\author [R.~Miranda, D.~Lai and H.~M\'{e}heut]
        {Ryan Miranda$^1$\thanks{rjm456@cornell.edu}, Dong Lai$^1$ and H\'{e}lo\"{i}se M\'{e}heut$^2$ \\
         $^1$Cornell Center for Astrophysics and Planetary Science, Department of Astronomy, Cornell University, Ithaca, NY 14853, USA \\
         $^2$Laboratoire AIM, CEA/DSM-CNRS-Universit\'{e} Paris 7, Irfu/Service d'Astrophysique, CEA-Saclay, F-91191 Gif-sur-Yvette, France}

\begin{document}

\maketitle

\begin{abstract}
The physical mechanism of angular momentum transport in poorly ionized regions of protoplanetary discs, the dead zones (DZs), is not understood. The presence of a DZ naturally leads to conditions susceptible to the Rossby wave instability (RWI), which produces vortices and spiral density waves that may revive the DZ and be responsible for observed large-scale disc structures. We present a series of two-dimensional hydrodynamic simulations to investigate the role of the RWI in DZs, including its impact on the long-term evolution of the disc and its morphology. The nonlinear RWI can generate Reynolds stresses (effective $\alpha$ parameter) as large as $0.01 - 0.05$ in the DZ, helping to sustain quasi-steady accretion throughout the disc. It also produces novel disc morphologies, including azimuthal asymmetries with $m = 1, 2$, and atypical vortex shapes. The angular momentum transport strength and morphology are most sensitive to two parameters: the radial extent of the DZ and the disc viscosity. The largest Reynolds stresses are produced when the radial extent of the DZ is less than its distance to the central star. Such narrow DZs lead to a single vortex or two coherent antipodal vortices in the quasi-steady state. The edges of wider DZs evolve separately, resulting in two independent vortices and reduced angular momentum transport efficiency. In either case, we find that, because of the Reynolds stresses generated by the nonlinear RWI, gravitational instability is unlikely to play a role in angular momentum transport across the DZ, unless the accretion rate is sufficiently high.
\end{abstract}

\begin{keywords}
accretion, accretion discs -- hydrodynamics -- instabilities -- protoplanetary discs
\end{keywords}

\section{Introduction}

The transport of angular momentum is a central problem in the evolution of protoplanetary discs. The magnetorotational instability (MRI; Balbus \& Hawley 1991, 1998) in magnetized Keplerian discs can sustain turbulence which effectively transports angular momentum, provided that the disc material is sufficiently ionized. This condition is thought to be met in the inner disc ($\lesssim 0.1 - 1.0$ AU), due to thermal ionization, and in the outer disc (a few to $\gtrsim 10$'s of AU), due to irradiation by high energy photons and cosmic rays. At intermediate radii, the disc is cold and and weakly ionized, producing an MRI-inactive ``dead zone'' (DZ) in which turbulence is significantly suppressed in the disc midplane, although thin surface layers may remain turbulent (Gammie 1996; see Armitage 2011 for a review). The spatial extent of the DZ and its level of residual turbulence depend on many factors. These include the ionization fraction, which is affected by the abundance of dust grains (e.g., Sano et al. 2000; Desch \& Turner 2015) and shielding of cosmic rays by stellar winds (Cleeves et al. 2013), the role of non-ideal MHD effects, such as Ohmic diffusion, ambipolar diffusion, and Hall drift (Bai \& Stone 2013; Bai 2013, 2014a, 2014b; Lesur et al. 2014; see Turner et al. 2014 for a review), and the direct interaction of the magnetized stellar wind with disc surface layers (Russo \& Thompson 2015).

In the absence of MRI-driven turbulence, one or more alternative angular momentum transport processes must operate in order to allow protoplanetary discs to evolve in accordance with their typical observed lifetimes and accretion rates (a few Myr and $10^{-8}\,M_\odot/\mathrm{yr}$, e.g., Hartmann et al. 1998; Haisch, Lada \& Lada 2001). Several such mechanisms have been proposed, such as gravitational instability (e.g., Lodato \& Rice 2004; Rafikov 2015), disc winds (e.g., Bai \& Stone 2013), baroclinic instability (e.g., Klahr \& Bodenheimer 2003), and vertical shear instability (e.g., Urpin \& Brandenburg 1998). These hydrodynamical instabilities may produce locally correlated velocity fluctuations, leading to enhanced effective viscosity (Balbus \& Papaloizou 1999). In these cases, the transport of angular momentum does not necessarily result from turbulence, but instead may be due to large-scale structures.

A promising mechanism for reviving the DZ is the Rossby wave instability (RWI), a global, non-axisymmetric instability which arises at ``bumps'' in Keplerian discs (Lovelace et al. 1999; Li et al. 2000; M\'{e}heut et al. 2010, 2012a, 2012b, 2013; see review by Lovelace \& Romanova 2014). More specifically, RWI is associated with narrow radial minima of vortensity (vorticity divided by surface density), and leads to growth of vortices, which then merge into a single vortex (Li et al. 2001). The anticyclonic rotation (i.e., opposite to the bulk disc rotation) of the resulting vortex prevents its destruction by Keplerian shear, allowing it to potentially have a long lifetime (e.g., Godon \& Livio 1999).

The sharp gradients in viscosity at DZ edges\footnote{The abrupt onset of MRI at a threshold resistivity level ensures that the gradient of effective viscosity is sharp, even when that of the underlying ionization/resistivity is not (Lyra et al. 2015).} naturally produce RWI-unstable vortensity profiles. Vortices and associated spiral density waves produced by the RWI create fluid stresses that transport angular momentum, which can revive the DZ (Varni\`{e}re \& Tagger 2006; Lyra et al. 2009b; Lyra \& Mac Low 2012; Reg\'{a}ly et al. 2012). Further, the low viscosity DZ is a favorable location for the production of vortices, since viscosity can inhibit the RWI (Lin 2014; Gholipour \& Nejad-Asghar 2014), and influence the long-term survival of vortices (Fu et al. 2014a; Zhu \& Stone 2014). In this paper, we examine the role of RWI and vortices in the evolution of discs with DZs, building on existing work by considering both the inner and outer DZ edges together (especially in the case when they are close to each other), and by focusing on the long-term, quasi-steady behavior of the disc.

Anticylonic vortices, such as those produced by RWI, may play a role in planet formation due to their ability to trap dust particles (Barge \& Sommeria 1995; Godon \& Livio 2000; Tanga et al. 1996; Lyra et al. 2009b; M\'{e}heut et al. 2012c; Lyra \& Lin 2013; Zhu \& Stone 2014), which may produce the conditions needed for the formation of planetesimals (see review by Chiang \& Youdin 2010). However, feedback on the gas by the accumulated dust particles may subsequently destroy the vortex (e.g., Chang \& Oishi 2010; M\'{e}heut et al. 2012c; Fu et al. 2014b). Recent observations of transitions discs (discs with central holes in dust emission, see review by Espaillat et al. 2014) show strong asymmetries in mm-dust emission (e.g., van der Marel et al. 2013; Casassus et al. 2013; Isella et al. 2013), which could possibly be explained by dust trapping in vortices. These vortices may be produced at DZ edges (e.g., Reg\'{a}ly et al. 2012), or at the edges of gaps opened by planets (e.g., Zhu \& Stone 2014). In the simulations presented in this paper, novel asymmetries are produced by the gas dynamics associated with the presence of a DZ. Since we do not follow the dynamics of dust particles, which are coupled to the gas through aerodynamic drag, we cannot make concrete predictions about the resultant dust morphologies. However, some qualitative features can be extrapolated to the behavior of dust, such as azimuthal symmetries, and the location of features relative to the position of the DZ.

It has been suggested that the presence of a DZ may lead to episodic accretion. In this scenario, accumulation of mass in the DZ triggers local gravitational instability driven turbulence, which heats the DZ and briefly triggers MRI, causing an accretion outburst, before cooling and repeating the cycle (e.g., Zhu et al. 2010a, 2010b; Martin \& Lubow 2011, 2014). This mechanism has been used to explain the outburst behavior of FU Orionis systems. The majority of work on this topic has used one-dimensional models (with some exceptions, e.g., Bae et al. 2014), which neglect non-axisymmetric effects. In this work, we show that, as long as the accretion rate is not too high, vortices produced by the RWI (which is explicitly non-axisymmetric) can generate fluid stresses which facilitate steady, non-episodic accretion through the DZ. We place a limit on the factor by which the gravitational stability parameter (Toomre $Q$) is reduced in the DZ compared to a steady-state disc with no DZ.

In this paper, we give considerable attention to narrow DZs, for which the radial extent of the DZ is smaller than its distance from the central star. As we show, this configuration leads to large fluid stresses and produces novel morphologies, including unusual vortex shapes and azimuthal symmetries (mode numbers other than $m = 1$). These features are the result of coherent oscillations of the entire DZ. By contrast, for wide DZs, each edge behaves independently, and the RWI evolves toward the familiar $m = 1$ symmetry (e.g., M\'{e}heut et al. 2012b), producing a much smaller Reynolds stress in the DZ.  Estimates of the extent of the DZ in realistic discs are highly uncertain, and range from very narrow, as described here, to highly extended. Viscosity profiles resembling our parameterized DZs (i.e., regions with reduced viscosity relative to their surroundings) may be present near or in between ice lines, where there are significant changes in dust grain abundance, on which the strength of MRI turbulence depends sensitively (e.g., Kretke \& Lin 2007; Bitsch et al. 2014). Therefore, there are plausible physical conditions in protoplanetary discs which may produce the DZ configurations modeled in this paper.

The main goal of this paper is to study the evolution of DZs using global, long-term hydrodynamic simulations. We find three main results. Discs with DZs evolve toward quasi-steady states, in which steady, non-episodic accretion is partially facilitated by fluid stresses produced by vortices and spiral density waves resulting from the RWI. We show that narrow DZs produce large Reynolds stresses, with only moderate surface density enhancements in the DZ, so they are unlikely to experience gravitational instability unless the accretion rate is high. We also present novel gas disc morphologies produced by the RWI in DZs.

The outline of this paper is as follows. In Section \ref{sec:setup}, we describe the the setup of our simulations. In Section \ref{sec:canonical_run}, we present a detailed analysis of a canonical run, which demonstrates our main results. We present a suite of simulations which systematically explore the effects of varying the DZ parameters in Section \ref{sec:parameter_study}. Finally, we discuss these results and conclude in Section \ref{sec:discussion}. In the Appendix, we present a linear calculation of how viscosity affects the RWI.

\section{Setup}
\label{sec:setup}
We consider a 2D (height-integrated) disc described by surface density $\Sigma$ and velocity $\boldsymbol{u} = \left(u_r, u_\phi\right)$, with a radial extent of $r_\mathrm{in} = 1$ to $r_\mathrm{out} = 12$. It experiences a gravitational potential (per unit mass) $\Phi = -GM_*/r$, where $M_*$ is the mass of the central star, so the Keplerian orbital frequency is $\Omega_\mathrm{K} = (GM_*/r^3)^{1/2}$. We adopt a locally isothermal equation of state $P = c_\mathrm{s}^2\left(r\right) \Sigma$, where $P$ is the height-integrated pressure, and
\be
c_\mathrm{s}\left(r\right) = c_{\mathrm{s}0} \left(\frac{r}{r_\mathrm{in}}\right)^{-1/2}
\ee
is the radially-dependent sound speed. The scale height of the disc, $H = c_\mathrm{s}/\Omega_\mathrm{K}$, is proportional to $r$, so that the aspect ratio $h = H/r$ is constant. The kinematic viscosity is prescribed by
\be
\nu = \alpha c_\mathrm{s} H.
\ee
The dimensionless viscosity parameter $\alpha$ is a function of $r$, given by
\be
\alpha = \alpha_0 f_\alpha\left(r\right),
\ee
where $f_\alpha\left(r\right)$, which describes the shape of the DZ, is given by
\be
f_\alpha\left(r\right) = 1 + \frac{1}{2}\left(1-\epsilon_\mathrm{DZ}\right)\left[\tanh\left(\frac{r-r_\mathrm{ODZ}}{\Delta r_\mathrm{ODZ}}\right) - \tanh\left(\frac{r-r_\mathrm{IDZ}}{\Delta r_\mathrm{IDZ}}\right)\right].
\ee
The meaning of the parameters are as follows: $r_\mathrm{IDZ}$ and $r_\mathrm{ODZ}$ are the inner and outer edges of the DZ, $\alpha_0$ is  the value of $\alpha$ in the active zones ($r < r_\mathrm{IDZ}$ and $r > r_\mathrm{ODZ}$), $\epsilon_\mathrm{DZ}$ is the factor by which $\alpha$ is reduced in the DZ (i.e., $\alpha = \epsilon_\mathrm{DZ}\alpha_0$ in the DZ), $\Delta r_\mathrm{IDZ}$ and $\Delta r_\mathrm{ODZ}$ are the widths of viscosity transitions at the inner and outer DZ edges. These, along with the aspect ratio $h$, fully describe each simulation. Unless otherwise stated, time is expressed in units of the Keplerian orbital period at $r_\mathrm{in}$, $P_\mathrm{in} = 2\pi/\Omega_\mathrm{in}$.

\subsection{Initial Conditions}
The initial surface density (in code units) is 
\be
\Sigma = r^{-1/2},
\ee
which satisfies the steady state condition $\Sigma \nu = \mathrm{constant}$ in the active zones ($\alpha = \alpha_0$), but not in the DZ, which would require $\Sigma$ to be larger by a factor of $\epsilon_\mathrm{DZ}^{-1}$. The disc is initially in centrifugal balance with $u_\phi = r \Omega_0$, where
\be
\label{eq:centrifugal}
\Omega^2_0 = \Omega^2_\mathrm{K} + \frac{c_\mathrm{s}^2}{r^2}\left(\frac{\mathrm{d} \ln \Sigma}{\mathrm{d} \ln r} + \frac{\mathrm{d} \ln c_\mathrm{s}^2}{\mathrm{d} \ln r}\right),
\ee
and initially $u_r = 0$ everywhere. Non-axisymmetric instabilities are seeded by random perturbations (i.e., they have no preferred azimuthal symmetry) to $\Sigma$, with amplitude $\delta\Sigma/\Sigma \lesssim 10^{-6}$.

\subsection{Boundary Conditions}
At the outer boundary ($r_\mathrm{out} = 12$), the fluid variables are fixed at the constant values
\be
\Sigma = r^{-1/2}, \quad u_r = -\frac{3}{2} \frac{\nu}{r}, \quad u_\phi = r \Omega_0,
\ee
so that the accretion rate,
\be
\dot{M} = -\int_0^{2\pi} \Sigma u_r r \mathrm{d} \phi,
\ee
supplied to the disc has the prescribed value $\dot{M}_0$. At the inner boundary ($r_\mathrm{in} = 1$), zero-gradient conditions are imposed on $\Sigma$ and $u_r$, while $u_\phi$ is fixed at its initial (modified) Keplerian value, i.e.,
\be
\frac{\partial \Sigma}{\partial r} = \frac{\partial u_r}{\partial r} = 0, \quad u_\phi = r \Omega_0.
\ee

Near the inner and outer boundaries, wave damping zones are used to minimize wave reflection. This is achieved by solving the following equations for the variables $\mathbf{x} = \left(\Sigma, u_r, u_\phi \right)$ at the end of each time step. In the outer damping zone $10 < r < 12$, they are damped to an equilibrium state,
\be
\label{eq:outer_bc}
\frac{\mathrm{d}\mathbf{x}}{\mathrm{d}t} = -\frac{\mathbf{x}-\mathbf{x}_0}{\tau_\mathrm{out}} R\left(r\right),
\ee
where $\mathbf{x}_0 = \left(r^{-1/2}, -3\nu/2r, r \Omega_0\right)$, so that the accretion rate is relaxed to $\dot{M}_0$. In the inner damping zone ($1 < r < 2$), they are damped to their azimuthal averages,
\be
\frac{\mathrm{d}\mathbf{x}}{\mathrm{d}t} = -\frac{\mathbf{x}-\langle \mathbf{x} \rangle_\phi}{\tau_\mathrm{in}} R\left(r\right).
\ee
We choose the damping timescales $\tau_\mathrm{in} = 2\pi\Omega_\mathrm{in}^{-1}$ and $\tau_\mathrm{out} = 2 \pi\Omega_\mathrm{out}^{-1}/10$, and the dimensionless envelope function is
\be
R\left(r\right) = 1 - \left(\frac{r-r_\mathrm{b}}{r_\mathrm{d}-r_\mathrm{b}}\right)^2,
\ee
where $r_\mathrm{b}$ is the (inner or outer) boundary and $r_\mathrm{d}$ is where the corresponding damping zone begins. We verified that our numerical results are not strongly affected by the placement of the boundaries and damping zones.

\subsection{Numerical Method}
The fluid equations are solved using the finite volume, Godunov scheme hydrodynamics code \textsc{pluto} (Mignone et al. 2007). We use second-order Runge-Kutta time stepping, linear spatial reconstruction, and a Roe method Riemann solver. Parabolic terms due to viscosity are handled using a super-time-stepping technique, and we utilize the FARGO advection algorithm, which relaxes the restrictive Courant condition associated with the average orbital motion of the disc. We use a static polar ($r, \phi$) grid with uniform $\Delta r$ and $\Delta\phi$, with a canonical resolution of $N_r \times N_\phi = 512 \times 256$.

\section{Canonical Run}
\label{sec:canonical_run}
The parameters of our canonical run (also referred to as ``run 00'' from here on) are as follows. The disc has an aspect ratio $h = 0.1$ and an active zone viscosity parameter $\alpha_0 = 0.1$. The DZ, in which the viscosity parameter is reduced by a factor $\epsilon_\mathrm{DZ} = 0.1$, extends from $r_\mathrm{IDZ} = 4.5$ to $r_\mathrm{ODZ} = 6.5$. The width of the viscosity transition at each DZ edge is equal to half of the local scale height, i.e., $\Delta r_\mathrm{IDZ} = hr_\mathrm{IDZ}/2 = 0.225$ and $\Delta r_\mathrm{ODZ} = hr_\mathrm{ODZ}/2 = 0.325$. The viscous timescale is $t_\mathrm{visc} \sim r^2/\nu \sim 160 (r/r_\mathrm{in})^{3/2}\,P_\mathrm{in}$, except in the DZ, where it is longer due to the reduction of $\alpha$, and near the DZ edges, where it is shorter due to sharp viscosity gradients. Therefore, we run the simulation for $10^4\,P_\mathrm{in}$, which captures several ($\sim 4$) viscous timescales near $r_\mathrm{ODZ}$, and more than one viscous timescale at $r_\mathrm{out}$.

\subsection{Evolution}

\begin{figure*}
\begin{center}
\includegraphics[width=0.749\textwidth,clip]{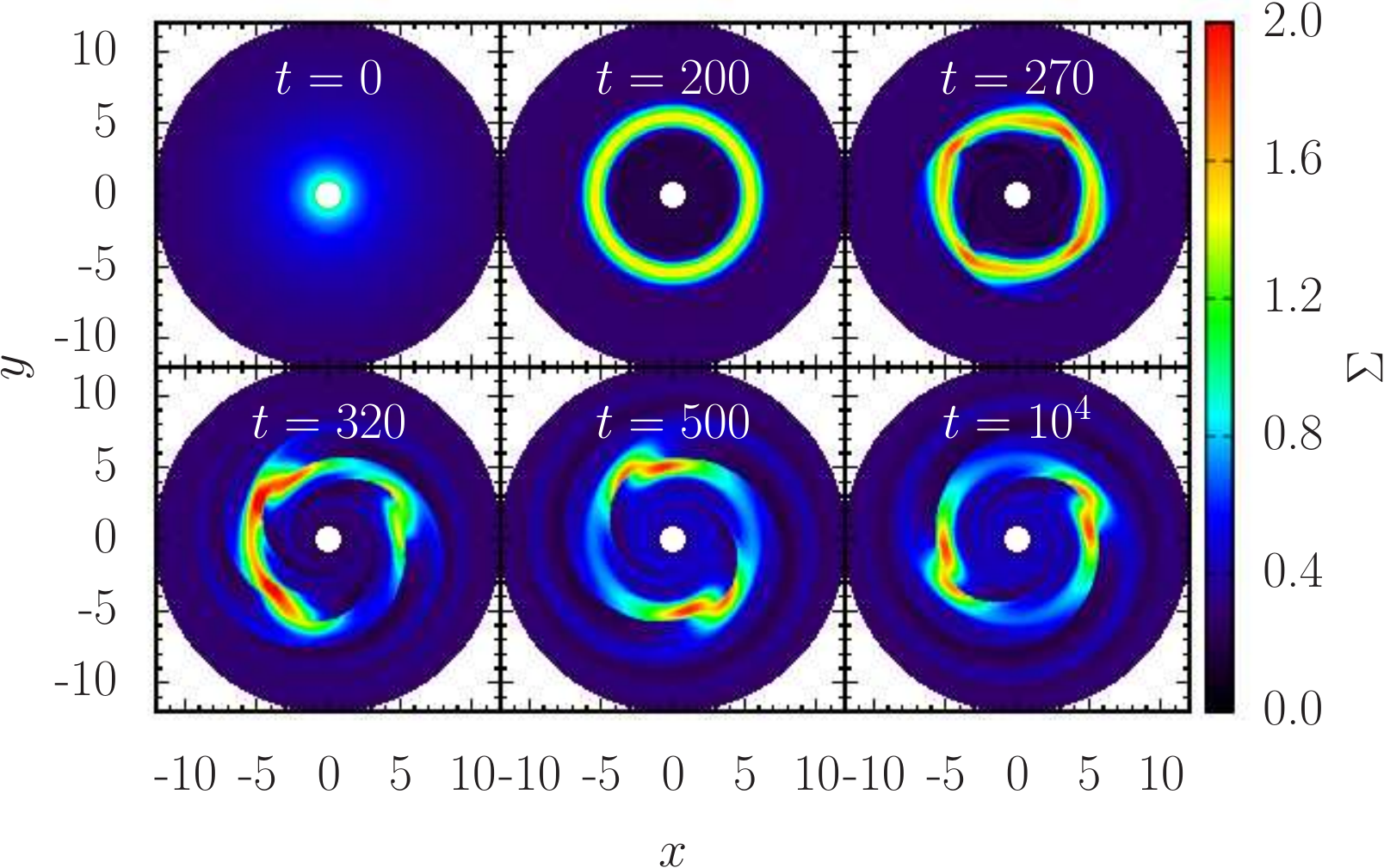}
\caption{Surface density at several times during the evolution of the canonical run. From top left to bottom right, the panels represent: (i) the initial condition, (ii) the axisymmetric bump before becoming unstable, (iii) the saturated state of the RWI with a prominent $m = 4$ pattern, (iv) the nonlinear saturation of the RWI and merging of the four vortices into three, (v) further vortex merging, resulting in two vortices, (vi) the quasi-steady state, with two persistent vortices.}
\label{fig:snapshots_canonical_evolution}
\end{center}
\end{figure*}

\begin{figure}
\begin{center}
\includegraphics[width=0.49\textwidth,clip]{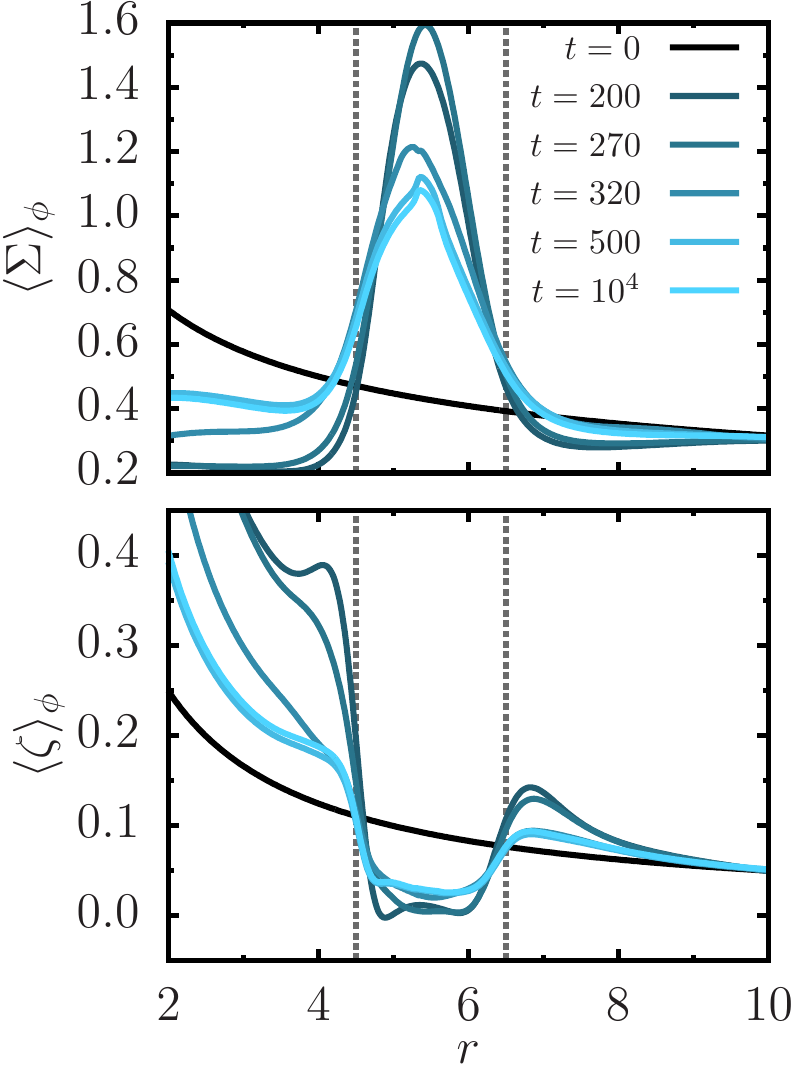}
\caption{Azimuthally-averaged surface density and vortensity profiles at the same times as shown in Fig. \ref{fig:snapshots_canonical_evolution}.}
\label{fig:rho_zeta_canonical_evolution}
\end{center}
\end{figure}

\begin{figure}
\begin{center}
\includegraphics[width=0.49\textwidth,clip]{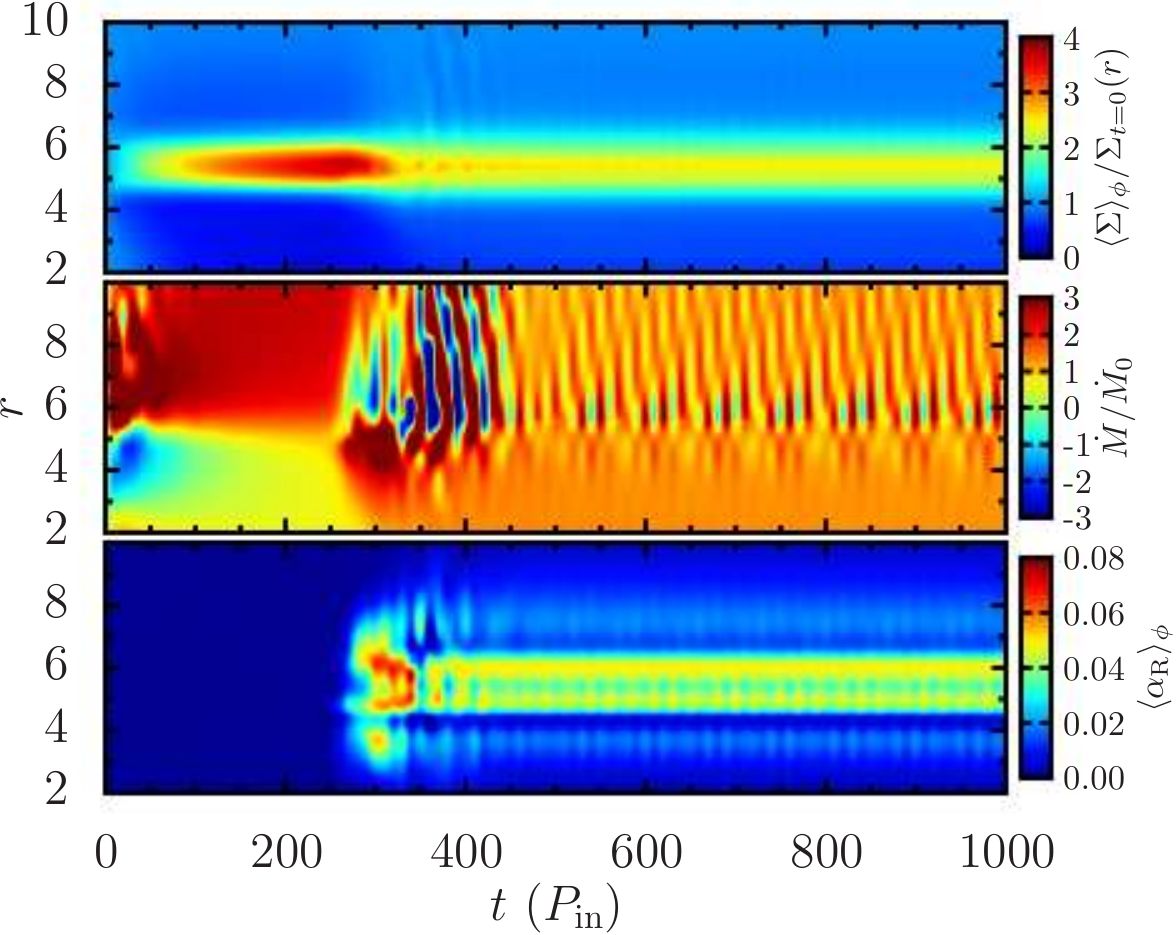}
\caption{Evolution of the canonical run for $t < 1000\,P_\mathrm{in}$. From top to bottom, the panels show the azimuthally-averaged surface density $\langle\Sigma\rangle_\phi$ (normalized by the initial profile), the accretion rate $\dot{M}$, and the azimuthally-averaged Reynolds stress, $\langle\alpha_\mathrm{R}\rangle_\phi$, as functions of $r$ and $t$.}
\label{fig:canonical_evolution}
\end{center}
\end{figure}

\begin{figure}
\begin{center}
\includegraphics[width=0.49\textwidth,clip]{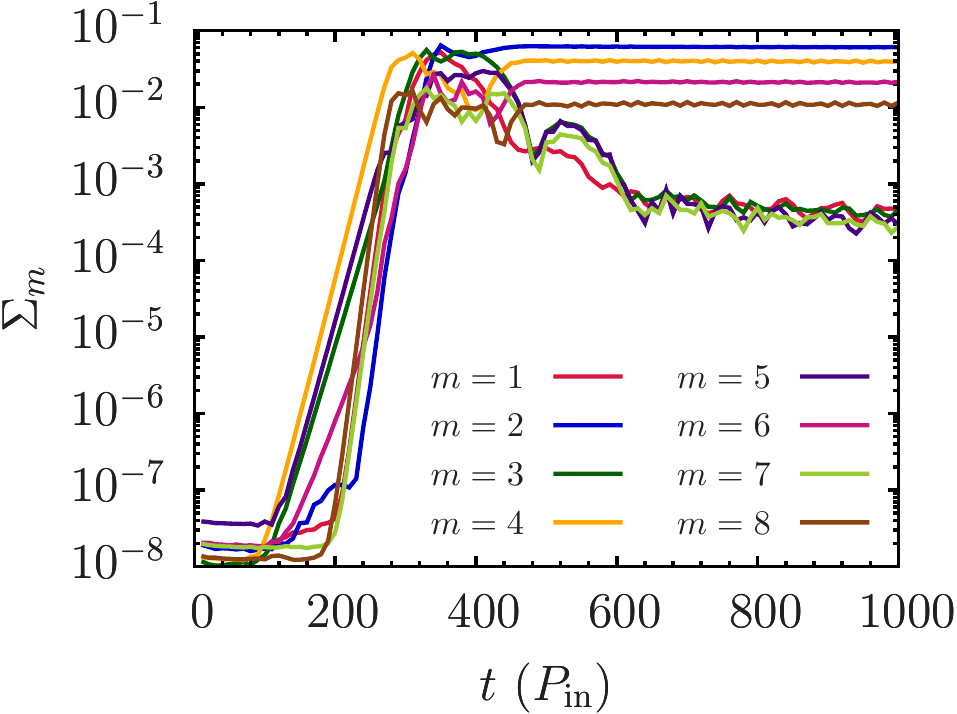}
\caption{Evolution of Fourier components of surface density, $\Sigma_m$ [Eq. (\ref{eq:sigma_m})], in the canonical run.}
\label{fig:m_evolution}
\end{center}
\end{figure}

The evolution of the canonical run is summarized in Fig. \ref{fig:snapshots_canonical_evolution}, which shows $\Sigma$ at six representative times. Initially, the surface density in the DZ increases, as a result of the accretion of mass from the outer disc stalling in the DZ. Consequently, an overdense ring forms in the DZ ($t = 200$), and the disc remains axisymmetric as the amplitude of the bump grows. Eventually, it becomes large enough to trigger RWI. Once it saturates ($t = 270$), the most prominent density perturbations have azimuthal number $m = 4$, i.e., there are four vortices. These then merge into three ($t = 320$), and finally two vortices ($t = 400$). The remaining two vortices, positioned $180^\circ$ apart in azimuth ($t = 500$), resist merging with one another, so the final morphology has an $m = 2$ symmetry, whose appearance changes only marginally over the remaining course of the simulation ($t = 10^4$). Azimuthally-averaged profiles of $\Sigma$ and vortensity $\zeta = |\nabla \times \mathbf{u}|/\Sigma$ at the same representative times are shown in Fig. \ref{fig:rho_zeta_canonical_evolution}. As the surface density in the DZ initially increases, forming a bump, RWI-unstable vortensity minima are produced. The peak of the density bump reaches a maximum that is $\sim 3$ times larger than the initial value before RWI is activated, which reduces the amplitude of the bump and smooths out the vortensity profile. By $t = 500$, both $\Sigma$ and $\zeta$ have nearly reached their steady-state profiles, and only evolve a small amount subsequently. In the final state, the peak $\langle\Sigma\rangle_\phi$ corresponds to a factor of two enhancement relative to the initial value.

Further details of the evolution are illustrated in Fig. \ref{fig:canonical_evolution}, which shows three azimuthally averaged quantities as functions of $r$ and $t$: surface density $\Sigma$ (normalized by the initial profile), accretion rate $\dot{M}$ (normalized by the supplied rate $\dot{M}_0$), and (dimensionless) Reynolds stress,
\be
\label{eq:reynolds}
\langle\alpha_\mathrm{R}\rangle_\phi = \frac{\langle\Sigma\tilde{u}_r \tilde{u}_\phi\rangle_\phi}{\langle\Sigma\rangle_\phi c_\mathrm{s}^2},
\ee
where $\tilde{u}_i = u_i - \langle u_i \rangle_\phi$. The development of a non-zero $\langle\alpha_\mathrm{R}\rangle_\phi$ at $t \approx 250$, which signifies enhanced angular momentum transport, is associated with the growth of the RWI, and is coincident with expulsion of mass from the DZ. This leads to a reduced peak $\langle\Sigma\rangle_\phi$, and a quasi-steady $\dot{M}$, which is modulated periodically, and with a large amplitude, by density waves, but has a steady time-averaged value close to $\dot{M}_0$ at all radii.

The growth, saturation and nonlinear evolution of the RWI are detailed in Fig. \ref{fig:m_evolution}, which shows the integrated Fourier components of $\Sigma$,
\be
\label{eq:sigma_m}
\Sigma_m\left(t\right) = \frac{1}{2\pi\left(r_\mathrm{out}-r_\mathrm{in}\right)} \int_{r_\mathrm{in}}^{r_\mathrm{out}} \int_0^{2\pi} \Sigma\left(r,\phi,t\right) \mathrm{e}^{-\mathrm{i} m \phi} \mathrm{d}\phi \mathrm{d}r,
\ee
as a function of time. Initially all components have approximately the same small amplitude ($\sim 10^{-8}$), since the seed perturbations have no preferred azimuthal number. They begin to grow exponentially after about $100\,P_\mathrm{in}$, when the density bump becomes RWI-unstable. Note that although there are two vortensity minima at this point (see Fig. \ref{fig:rho_zeta_canonical_evolution}), they do not become unstable independent of each another. Instead, their proximity allows them to interact with one another, so that the entire DZ is effectively a single site for RWI to occur. During the growth phase, all of the $\Sigma_m$'s have similar growth rates, but the $m = 4$ component grows fastest and is the first to saturate, with an amplitude of $\sim 10^{-2}$. By about $360\,P_\mathrm{in}$, the other components have saturated at comparable values. Energy is then transferred from high-$m$ to low-$m$ modes through vortex merging (see Fig. \ref{fig:snapshots_canonical_evolution}). In the final state ($t > 600$), $\Sigma_2$ is largest, followed by $\Sigma_4, \Sigma_6,$ and $\Sigma_8$, each of which is smaller than the previous by a factor of a few. The amplitudes of odd-$m$ components are several orders of magnitude smaller. This configuration can be interpreted as a nonlinear $m = 2$ mode.

\subsection{Quasi-Steady State}

\begin{figure}
\begin{center}
\includegraphics[width=0.49\textwidth,clip]{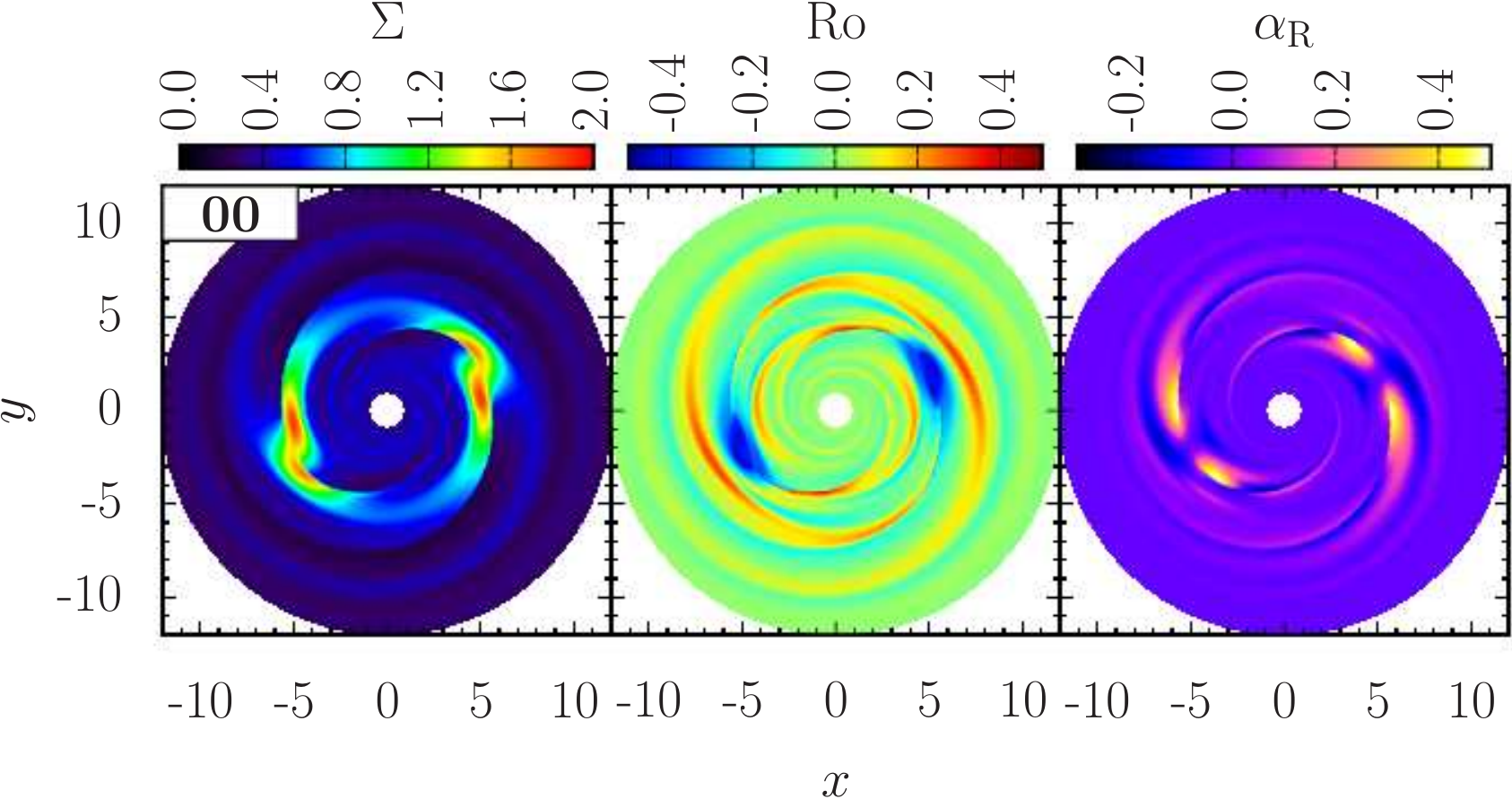}
\caption{Snapshot of surface density $\Sigma$, Rossby number $\mathrm{Ro}$, and Reynolds stress $\alpha_\mathrm{R}$ in the quasi-steady state of the canonical run.}
\label{fig:snapshot_canonical_steady}
\end{center}
\end{figure}

\begin{figure}
\begin{center}
\includegraphics[width=0.49\textwidth,clip]{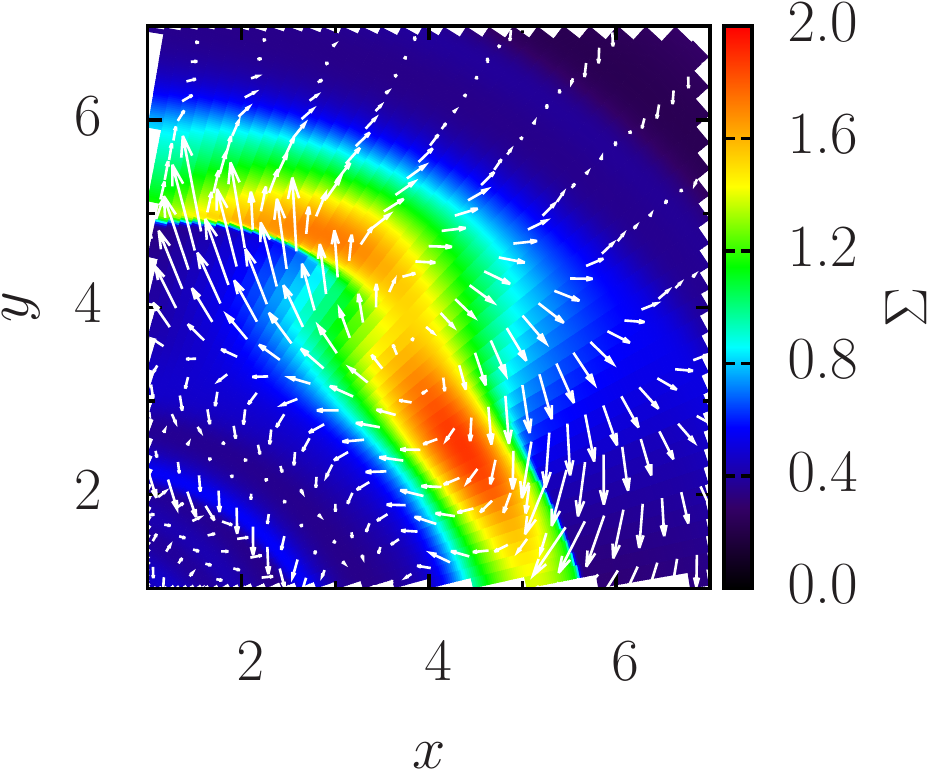}
\caption{Zoom-in of one of the vortices in the canonical run (there is another, nearly identical vortex on the opposite side of the disc), showing surface density (colors) and velocity (with the Keplerian velocity subtracted; arrows). The longest arrows represent velocities approximately $1.8$ times the local sound speed. The direction of the bulk disc rotation is anti-clockwise, so the clockwise motion of the velocity perturbations indicates that the vortex is anticylonic.}
\label{fig:vortex}
\end{center}
\end{figure}

After about $600\,P_\mathrm{in}$, the disc reaches a global quasi-steady state. In this stage, the azimuthally-averaged profiles of $\Sigma$ and $\alpha_\mathrm{R}$ remain steady, as illustrated by Fig. \ref{fig:canonical_evolution}. The accretion rate varies periodically in much of the disc due to waves launched from the DZ, but is steady in a time-averaged sense. The waves are damped in the inner disc ($r \lesssim 4$), resulting in almost steady accretion through the inner boundary. Further, the azimuthal structure of the disc (i.e., the amplitude of density perturbations proportional to $\mathrm{e}^{\mathrm{i}m\phi}$) does not evolve further (see Fig. \ref{fig:m_evolution}). This azimuthal structure is of interest because it is dominated by an $m = 2$ pattern. This differs from the morphology resulting from an isolated density bump in an inviscid disc (e.g., Meheut et al. 2012b), in which all of the initial RWI vortices merge into a single vortex, resulting in a global $m = 1$ symmetry. The two vortices resist merging for the entire $10^4\,P_\mathrm{in}$ duration of the simulation, which is much longer than the viscous timescale at the outer edge of the DZ. This morphology also appears in several other runs in our parameter study (see Section \ref{sec:parameter_study}).

The global structure of the disc in the quasi-steady state is illustrated in Fig. \ref{fig:snapshot_canonical_steady}, which shows snapshots of surface density, Rossby number, and local (dimensionless) Reynolds stress $\alpha_\mathrm{R} = \tilde{u}_r\tilde{u}_\phi/c_\mathrm{s}^2$. Here the Rossby number is defined
\be
\mathrm{Ro} = \frac{\tilde{\omega}}{2 \Omega_\mathrm{K}},
\label{eq:rossby}
\ee
where $\tilde{\omega} = \left[\nabla \times \left(\boldsymbol{u} - r \Omega_\mathrm{K} \hat{\boldsymbol{\phi}} \right)\right]_z$ is the residual (total minus Keplerian) vorticity. The nonlinear $m = 2$ morphology is apparent in the snapshots of all three quantities. The surface density features associated with this azimuthal symmetry are coincident with regions of approximately constant $\mathrm{Ro}$, which is characteristic of vortices. They have $\mathrm{Ro} \approx -0.5$, or $\tilde{\omega} \approx -\Omega_\mathrm{K}$, which means that they approximately rotate with the local shear (for a Keplerian flow, the local shear is $-3\Omega_\mathrm{K}/2$), and thus their survival is not threatened by the shear. The center of the vortices have small Reynolds stresses, while large positive stress is produced at their edges. The angular momentum transport associated with the vortices and the associated spiral density waves outside of the DZ is described by the azimuthally-averaged Reynolds stress (see Fig. \ref{fig:canonical_evolution}), which can be interpreted as an effective viscosity parameter. It is peaked at about $0.05$ (near the outer DZ edge), but remains larger than $0.03$ even in the center of the DZ. Since this is larger than the intrinsic DZ viscosity, $\epsilon_\mathrm{DZ}\alpha_0 = 0.01$, it plays a significant role in maintaing steady accretion through the DZ.

Figure \ref{fig:vortex} gives a detailed view of one of the two persistent vortices in the quasi-steady state. It consists of two dense blobs which rotate anticyclonically about a slightly less dense core (although the core has a surface density several times larger than the surrounding disc). The shape of the vortex is notable, as it differs significantly from that of a typical RWI vortex, which consists of a smoother surface density profile with only a single maximum. Although the overdense feature is large, with a radial extent similar to that of the DZ itself, and spanning nearly $90^\circ$ in azimuth, the vortex proper, i.e., the region of negative vorticity (see Fig. \ref{fig:snapshot_canonical_steady}) is smaller, with $\Delta r \approx 1.5$ and $r\Delta\phi \approx 4.0$. The largest velocity perturbations associated with it have an amplitude $v_\mathrm{max} \approx 1.8 \, c_\mathrm{s}$ (with $c_\mathrm{s}$ evaluated in the middle of the DZ). The radial size of the vortex is roughly consistent with $|v_\mathrm{max}/\tilde{\omega}| \approx 1.8H$ (where $\tilde{\omega} \approx -\Omega_\mathrm{K}$), the distance over which it can remain coherent given its rotation frequency and velocity. While the average surface density in the DZ is about two times larger than it would be in a constant $\alpha$ disc, the maximum surface density enhancement (in the centers of the blobs) is larger by a further factor of two.

\subsection{Periodicity}

\begin{figure}
\begin{center}
\includegraphics[width=0.49\textwidth,clip]{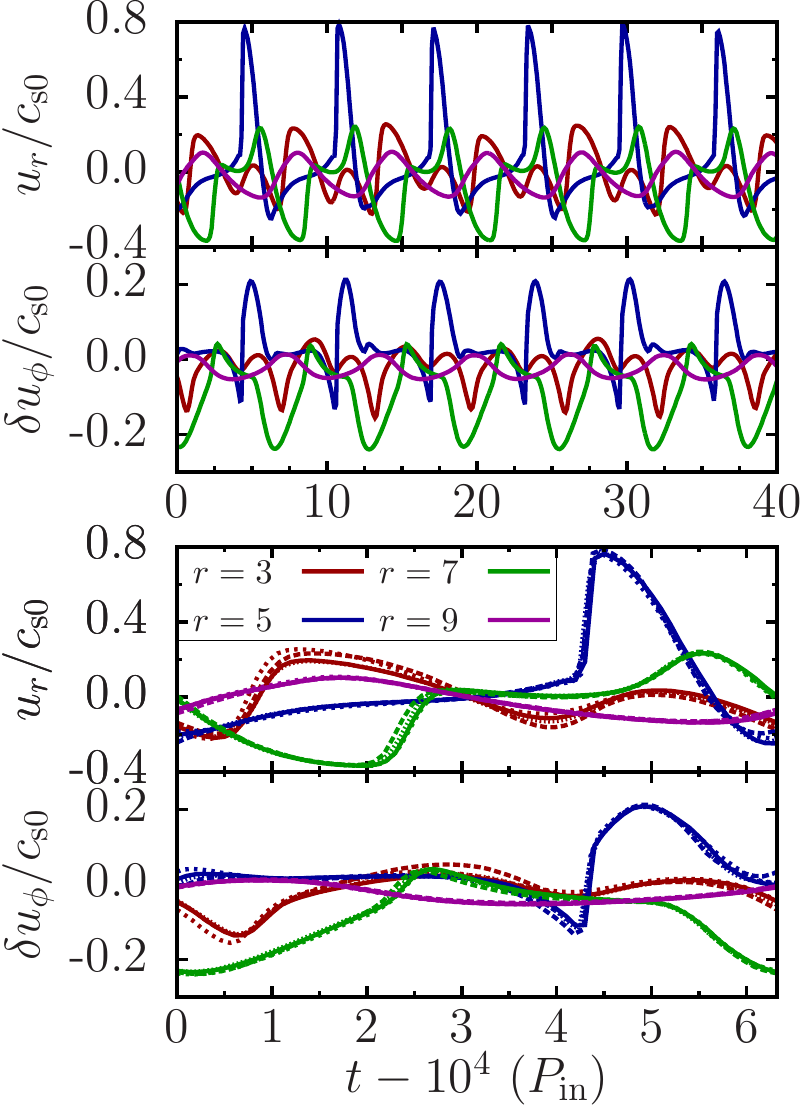}
\caption{Radial and azimuthal (minus Keplerian) velocities at several locations, specified by $(r,\phi) = (r,0)$, as a function of time during the quasi-steady state of the canonical run. The top panel shows several oscillation periods, and the bottom panel shows four of them (represented by different line types, e.g., solid, dashed, etc.) folded with a period of $6.31\,P_\mathrm{in}$.}
\label{fig:ur_uphi}
\end{center}
\end{figure}

A notable feature of the canonical run (as well as other runs in our parameter study) is that the velocity field of the disc is globally periodic, despite the presence of highly nonlinear waves (see Fig. \ref{fig:ur_uphi}). Figure \ref{fig:psd} (top left panel) shows the power spectrum of radial velocity $u_r$, i.e., the square of its temporal Fourier transform,
\be
\mathcal{F}\left[u_r\right]\left(r,\omega\right) = \int_{t_1}^{t_2} u_r\left(r,0,t\right) \mathrm{e}^{\mathrm{i}\omega t} \mathrm{dt}, 
\ee
where $t_1$ and $t_2$ are in the quasi-steady state of the disc with $t_2-t_1 = 100\,P_\mathrm{in}$. Power is concentrated in a global mode with frequency $\omega = 0.159\,\Omega_\mathrm{in}$ (and harmonics of this fundamental frequency). Delineating the power spectrum by azimuthal number reveals that it is an $m = 2$ mode, so this frequency is twice the pattern frequency of the mode. The pattern frequency, $\omega_\mathrm{P} = 0.079\,\Omega_\mathrm{in}$ is very close to the Keplerian frequency at the peak of the azimuthally-averaged surface density profile (located at $r_\mathrm{peak} = 5.38$), $\Omega_{\mathrm{K,peak}} = 0.080\,\Omega_\mathrm{in}$. This is characteristic of RWI modes, whose frequencies are $m$-times the corotation frequency of the unstable bump, but it is remarkable that this coherent global frequency persists in its nonlinear stage.

The global oscillations are quasi-periodic, and the power is spread around the peak frequency with a full-width half-maximum of $\Delta\omega/\omega = 8.5\%$. Thus, the velocity fluctuations at any (Eulerian) point, sampled one period apart in time, are not exactly equal to one another. Figure \ref{fig:ur_uphi} depicts the velocities $u_r$ and $u_\phi$ at several radii (evaluated at $\phi = 0$) as a function of time for several oscillation periods after the main $10^4\,P_\mathrm{in}$ run. The same periodicity is seen at all locations, since they are the result of a global mode. The lower panel gives a closer look at the quasi-periodicity by folding over four periods (with a period extracted from the power spectrum), demonstrating the slight variation between successive oscillation periods.

\subsection{Resolution and Convergence}
We performed the canonical run with double the number of radial and azimuthal grid points ($N_r \times N_\phi = 1024\times512$; we denote this run ``00\_res'') to determine how our our results depend on resolution. No qualitative differences are observed in the disc evolution (e.g., development of bump, growth of RWI), or the morphological appearance in the quasi-steady state ($m = 2$ symmetry and two-lobed vortex shape). The key numerical quantities characterizing the quasi-steady are given in Table \ref{tab:summary}, alongside those of the standard canonical run. They differ by less than $2\%$, which indicates that they are converged with respect to resolution in the standard run. We also performed the canonical run with a steeper initial surface density profile ($\Sigma \propto r^{-1}$), and found that it does not impact the final disc configuration, only the transient initial growth of the bump and the RWI. Therefore, the quasi-steady state that is eventually reached is not sensitive to the initial conditions, and only depends on the DZ geometry.

\section{Parameter Study}
\label{sec:parameter_study}

\begin{table*}
\begin{center}
\begin{tabular}{c|c|c|c|c|c|c|c|c|c|c|c|}
\hline

Run & $h$ & $\alpha_0$ & $\epsilon_\mathrm{DZ}$ & $r_\mathrm{IDZ}, r_\mathrm{ODZ}$ & $\Delta r_\mathrm{IDZ/ODZ}$ & $M_\mathrm{DZ}/M_\mathrm{nodz}$ & $\Sigma_\mathrm{max}/\Sigma_{0,\mathrm{dz}}$ & $\langle\langle\alpha_\mathrm{R}\rangle\rangle_{\phi,t,\mathrm{max}}$ & $\langle\langle\langle\alpha_\mathrm{R}\rangle\rangle_{\phi,t}\rangle_\mathrm{DZ}$ & $m$ \\ \hline

00      & $0.1$   & $0.1$   & $0.1$  & $4.5, 6.5$ & $H/2$  & $1.96$ & $4.34$  & $4.89\times10^{-2}$ & $3.97\times10^{-2}$ & $2$   \\
00\_res & $0.1$   & $0.1$   & $0.1$  & $4.5, 6.5$ & $H/2$  & $1.93$ & $4.26$  & $4.89\times10^{-2}$ & $4.03\times10^{-2}$ & $2$   \\ \hline
R1      & $0.1$   & $0.1$   & $0.1$  & $4.0, 7.0$ & $H/2$  & $4.71$ & $8.71$  & $1.89\times10^{-2}$ & $8.62\times10^{-3}$ & $2/2$ \\
R2      & $0.1$   & $0.1$   & $0.1$  & $3.5, 7.5$ & $H/2$  & $5.90$ & $11.10$ & $1.34\times10^{-2}$ & $4.10\times10^{-3}$ & $1/2$ \\
R3      & $0.1$   & $0.1$   & $0.1$  & $2.5, 8.5$ & $H/2$  & $6.50$ & $11.38$ & $8.92\times10^{-3}$ & $2.19\times10^{-3}$ & $1/1$ \\ \hline
DR1     & $0.1$   & $0.1$   & $0.1$  & $4.5, 6.5$ & $H$    & $2.09$ & $4.11$  & $2.58\times10^{-2}$ & $2.33\times10^{-2}$ & $2$   \\
DR2     & $0.1$   & $0.1$   & $0.1$  & $4.5, 6.5$ & $3H/2$ & $2.41$ & $3.66$  & $1.53\times10^{-3}$ & $5.85\times10^{-4}$ & $2$   \\ \hline
E1      & $0.1$   & $0.1$   & $0.01$ & $4.5, 6.5$ & $H/2$  & $2.05$ & $5.07$  & $6.00\times10^{-2}$ & $5.03\times10^{-2}$ & $2$   \\ \hline
C1      & $0.075$ & $0.1$   & $0.1$  & $4.5, 6.5$ & $H/2$  & $3.40$ & $14.20$ & $1.71\times10^{-2}$ & $1.27\times10^{-2}$ & $1$   \\
C2      & $0.05$  & $0.1$   & $0.1$  & $4.5, 6.5$ & $H/2$  & $6.21$ & $11.31$ & $1.05\times10^{-2}$ & $3.52\times10^{-3}$ & $3/2$ \\  \hline
A1      & $0.1$   & $0.05$  & $0.1$  & $4.5, 6.5$ & $H/2$  & $1.65$ & $3.47$  & $3.16\times10^{-2}$ & $2.63\times10^{-2}$ & $2$   \\
A2      & $0.1$   & $0.025$ & $0.1$  & $4.5, 6.5$ & $H/2$  & $1.81$ & $3.69$  & $1.42\times10^{-2}$ & $1.14\times10^{-2}$ & $1$   \\

\hline

\end{tabular}
\end{center}
\caption{Parameters and main quantitative results of our simulations. The first column gives the alphanumeric label of each run. The next five columns give the parameters of the runs: disc aspect ratio $h = H/r$, active zone viscosity parameter $\alpha_0$, DZ viscosity reduction factor $\epsilon_\mathrm{DZ}$, DZ edges $r_\mathrm{IDZ}$ and $r_\mathrm{ODZ}$, and viscosity transition widths $\Delta r_\mathrm{IDZ}$ and $r_\mathrm{ODZ}$ (see Section \ref{sec:setup}). The next column gives the mass in the DZ, $M_\mathrm{DZ}$, in the quasi-steady state, normalized by the mass at $t = 0$. The next column is the maximum surface density $\Sigma_\mathrm{max}$, relative to the average density in the middle of the DZ for a constant $\alpha$ disc. The next two columns give the maximum of the azimuthally-averaged and time-averaged (over $100\,P_\mathrm{in}$ in the quasi-steady state) dimensionless Reynolds stress, $\langle\langle\alpha_\mathrm{R}\rangle\rangle_{\phi,t,\mathrm{max}}$, and its average in the DZ, $\langle\langle\langle\alpha_\mathrm{R}\rangle\rangle_{\phi,t}\rangle_\mathrm{DZ}$. The last column describes the morphology of the quasi-steady state in terms of the dominant azimuthal mode number $m$. A single value in this column indicates the azimuthal number of a single global mode, which is coherent across the entire DZ. Two values separated by a slash indicate the $m$'s of two modes, localized to the outer (first value) and inner (second value) DZ edges.}
\label{tab:summary}
\end{table*}

\begin{figure*}
\begin{center}
\includegraphics[width=0.45\textwidth,clip]{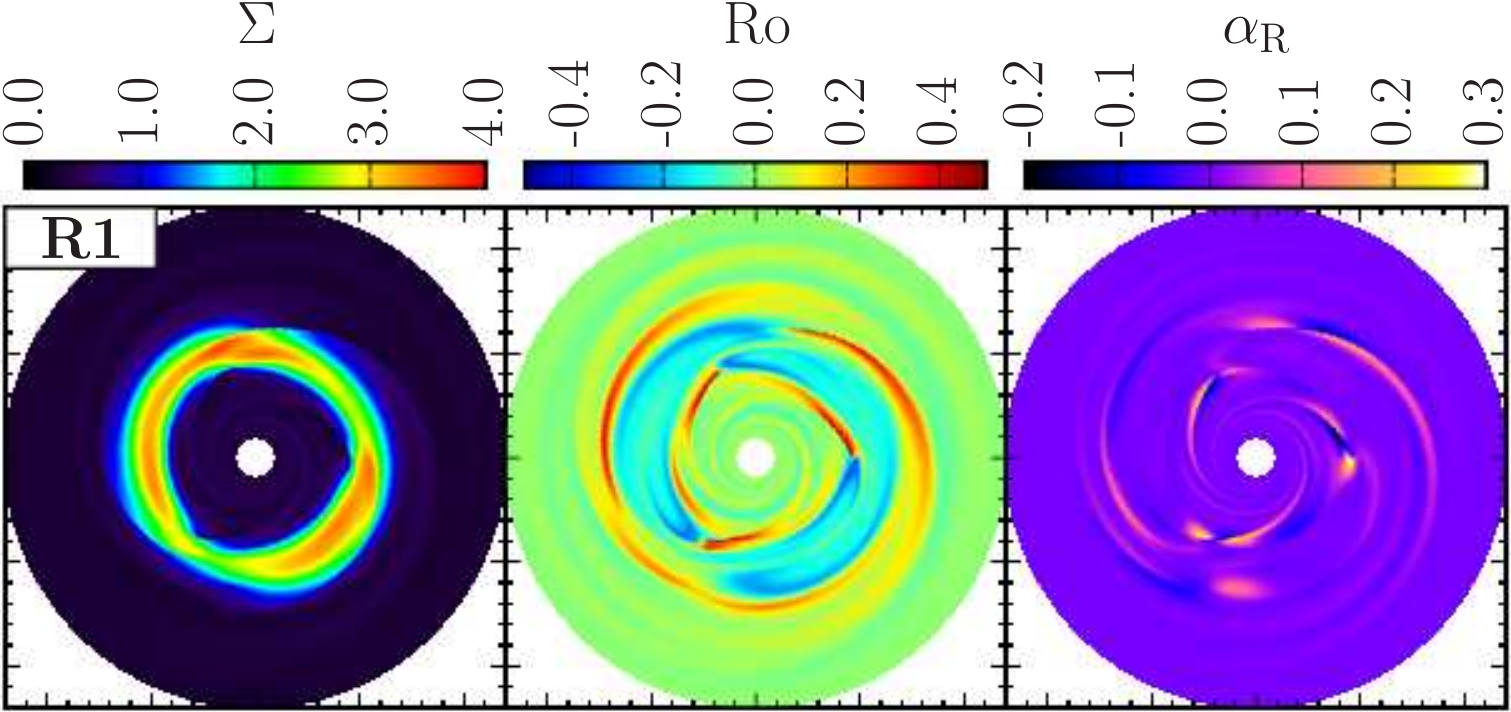}
\includegraphics[width=0.45\textwidth,clip]{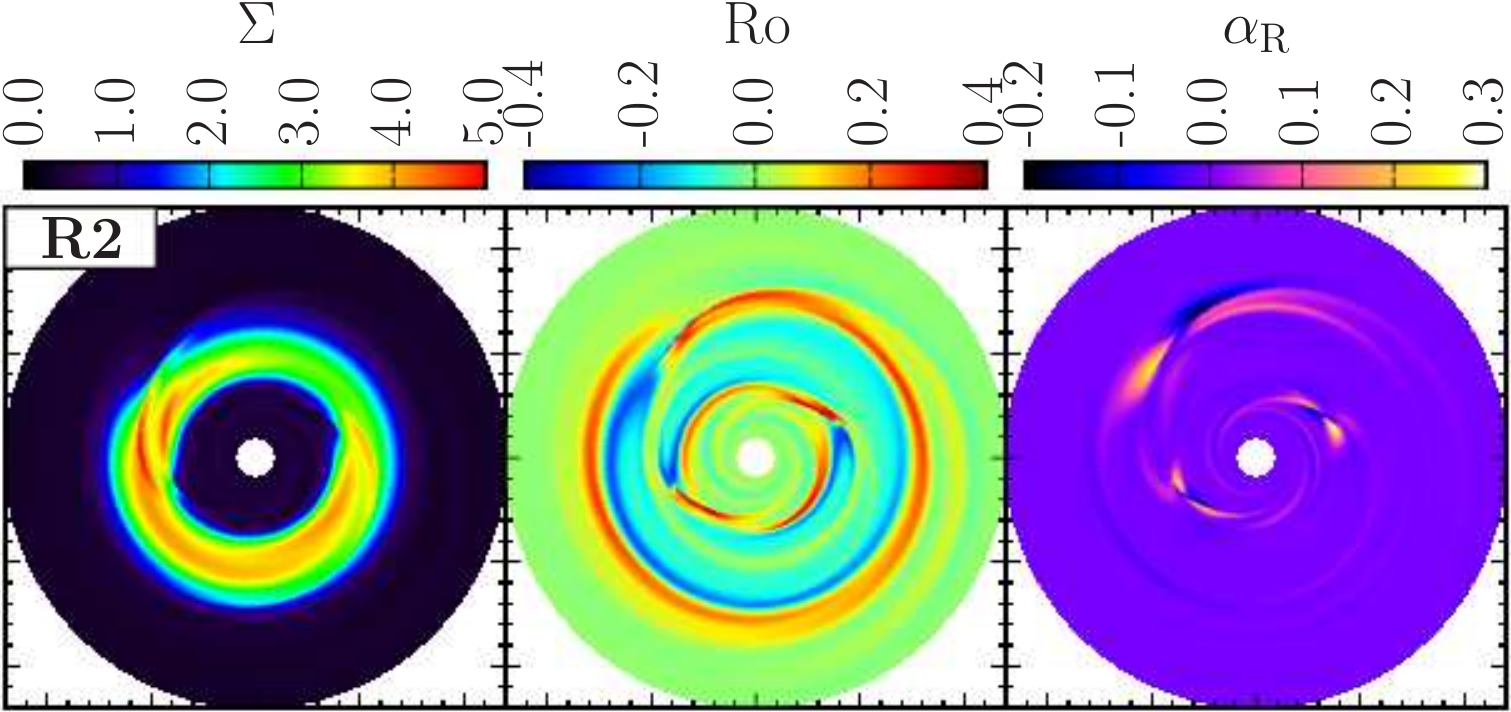}
\includegraphics[width=0.45\textwidth,clip]{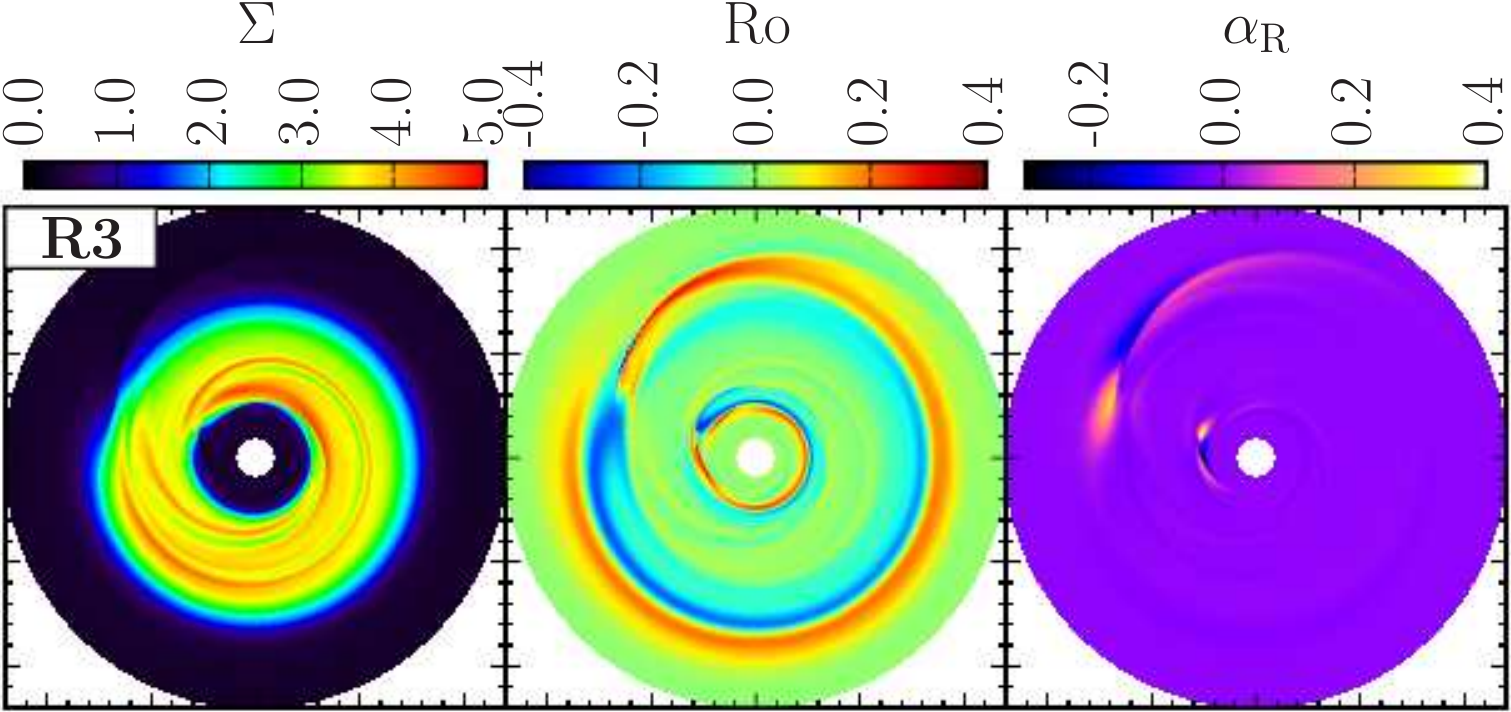}
\includegraphics[width=0.45\textwidth,clip]{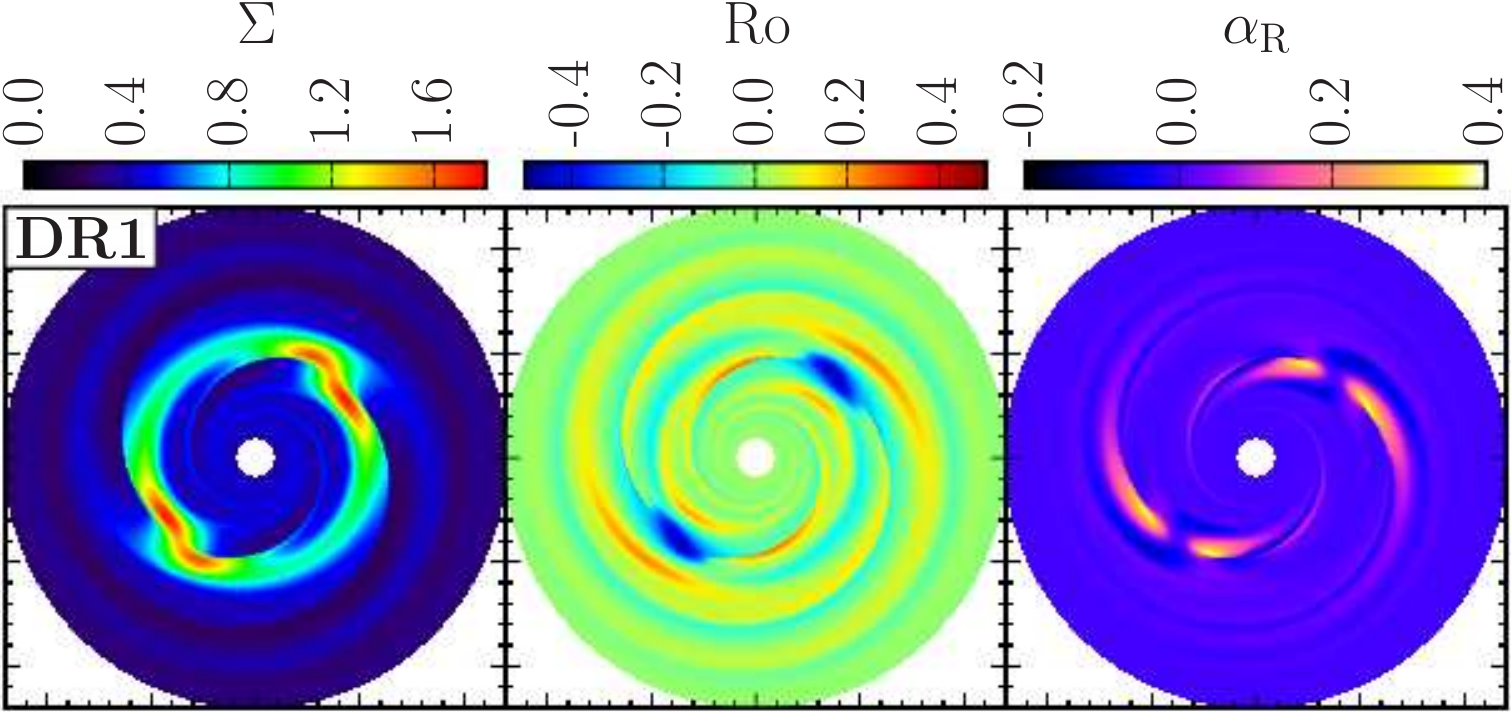}
\includegraphics[width=0.45\textwidth,clip]{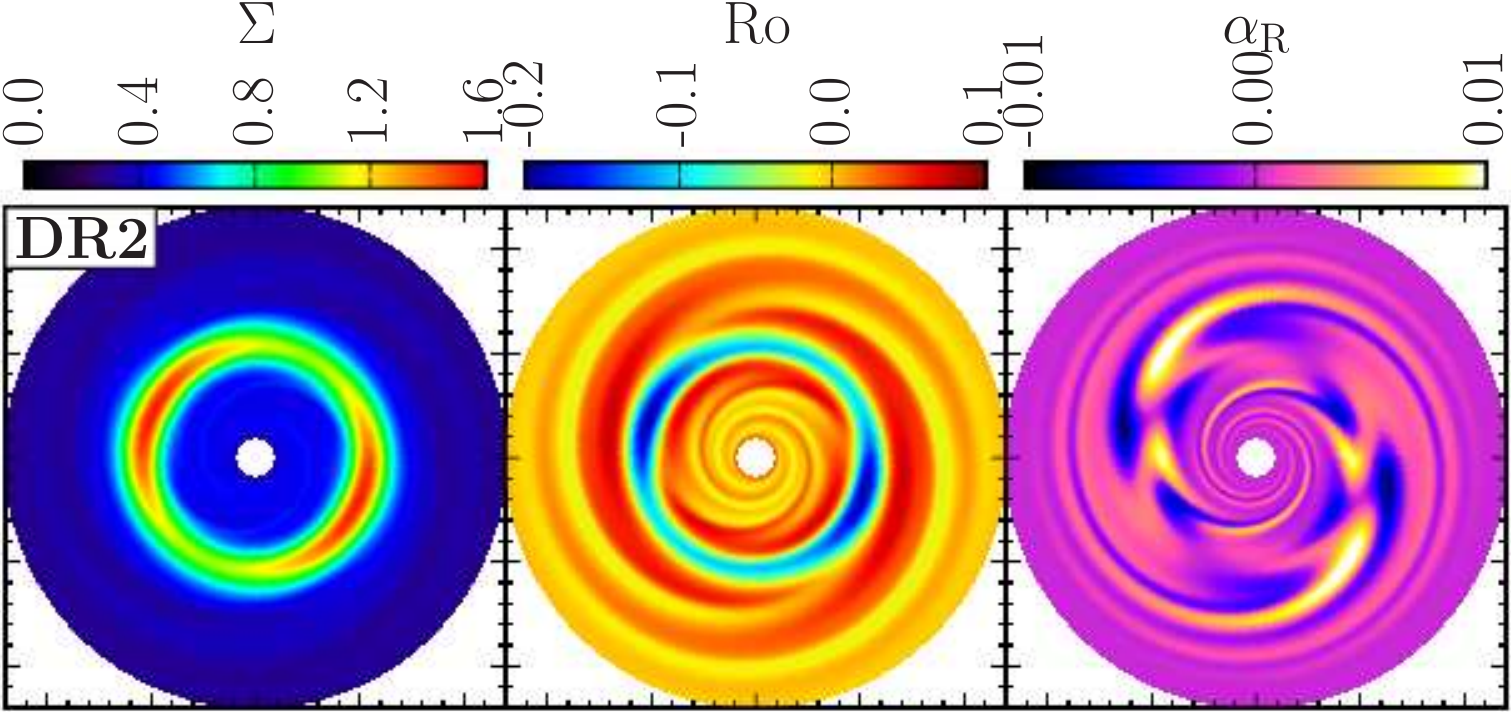}
\includegraphics[width=0.45\textwidth,clip]{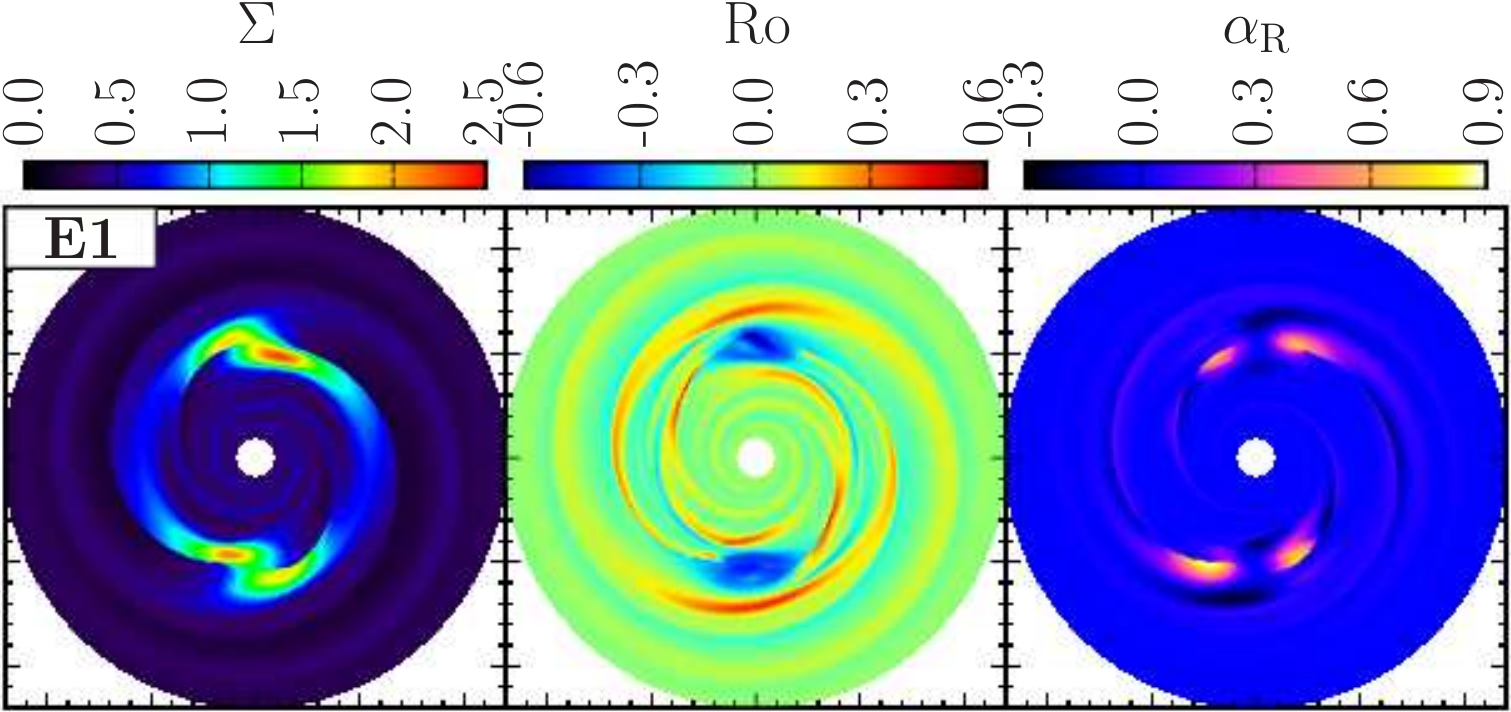}
\includegraphics[width=0.45\textwidth,clip]{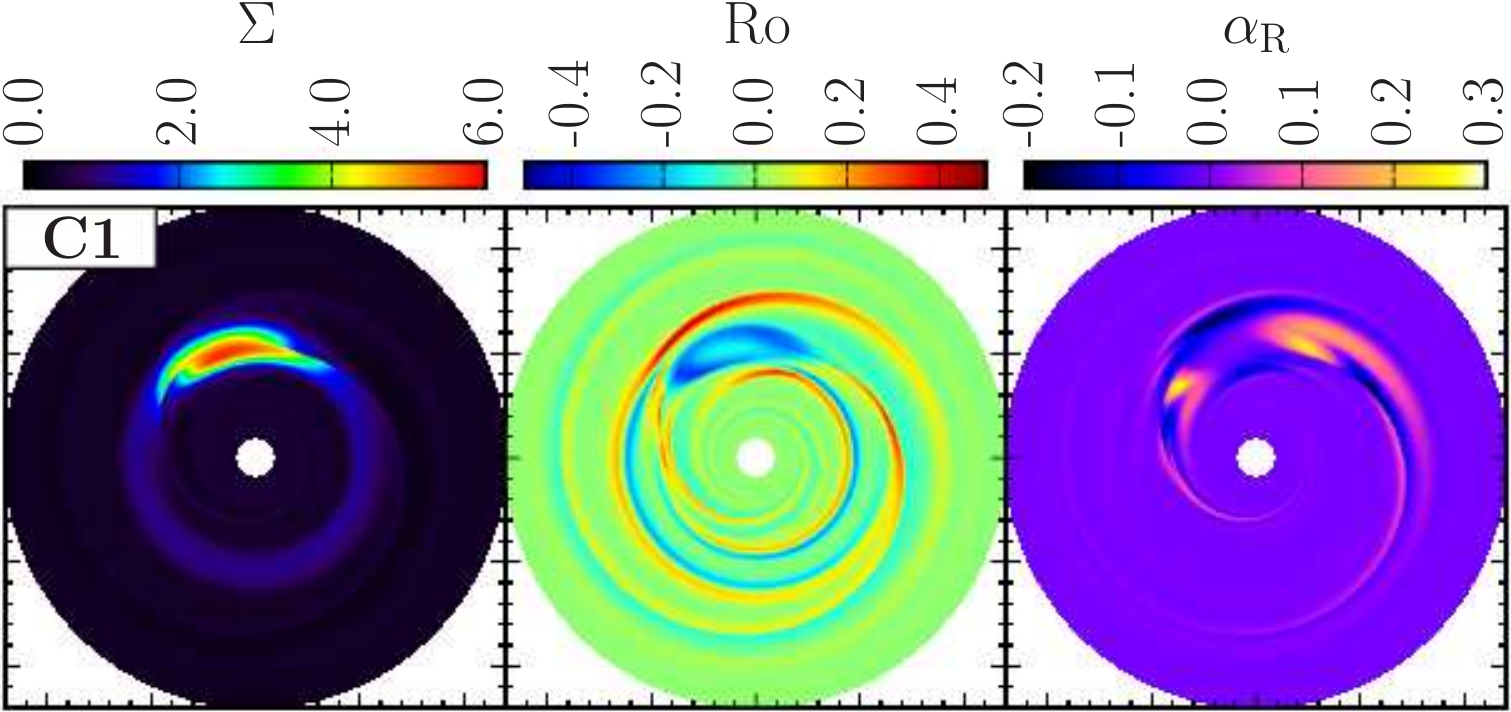}
\includegraphics[width=0.45\textwidth,clip]{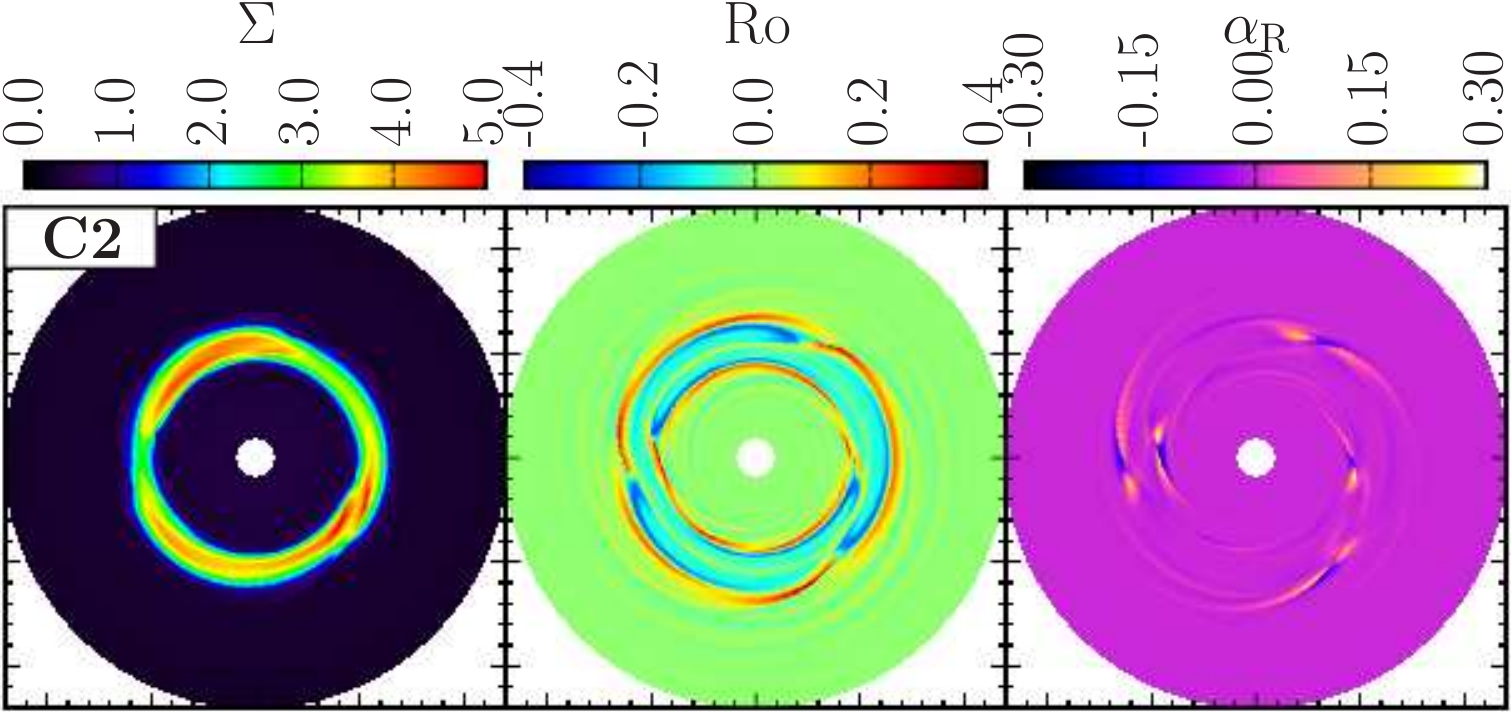}
\includegraphics[width=0.45\textwidth,clip]{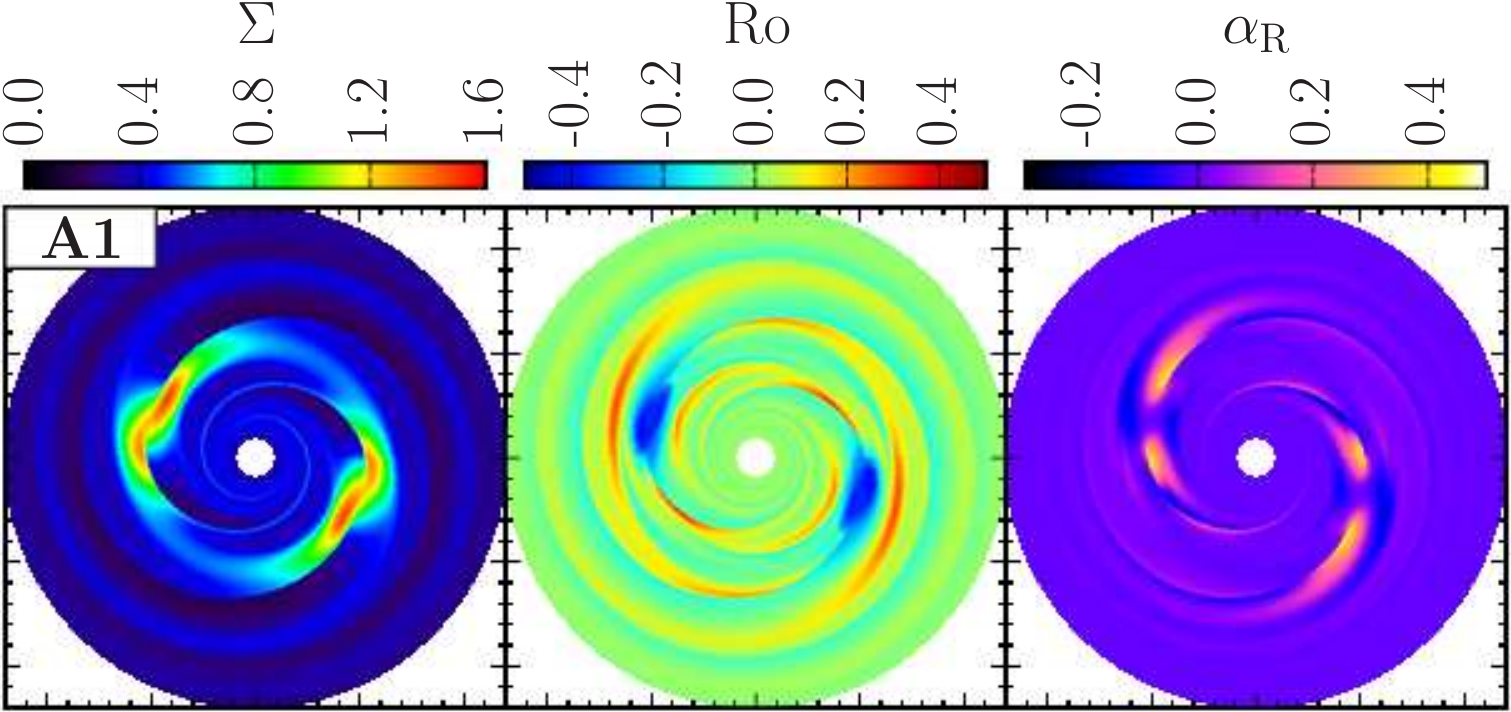}
\includegraphics[width=0.45\textwidth,clip]{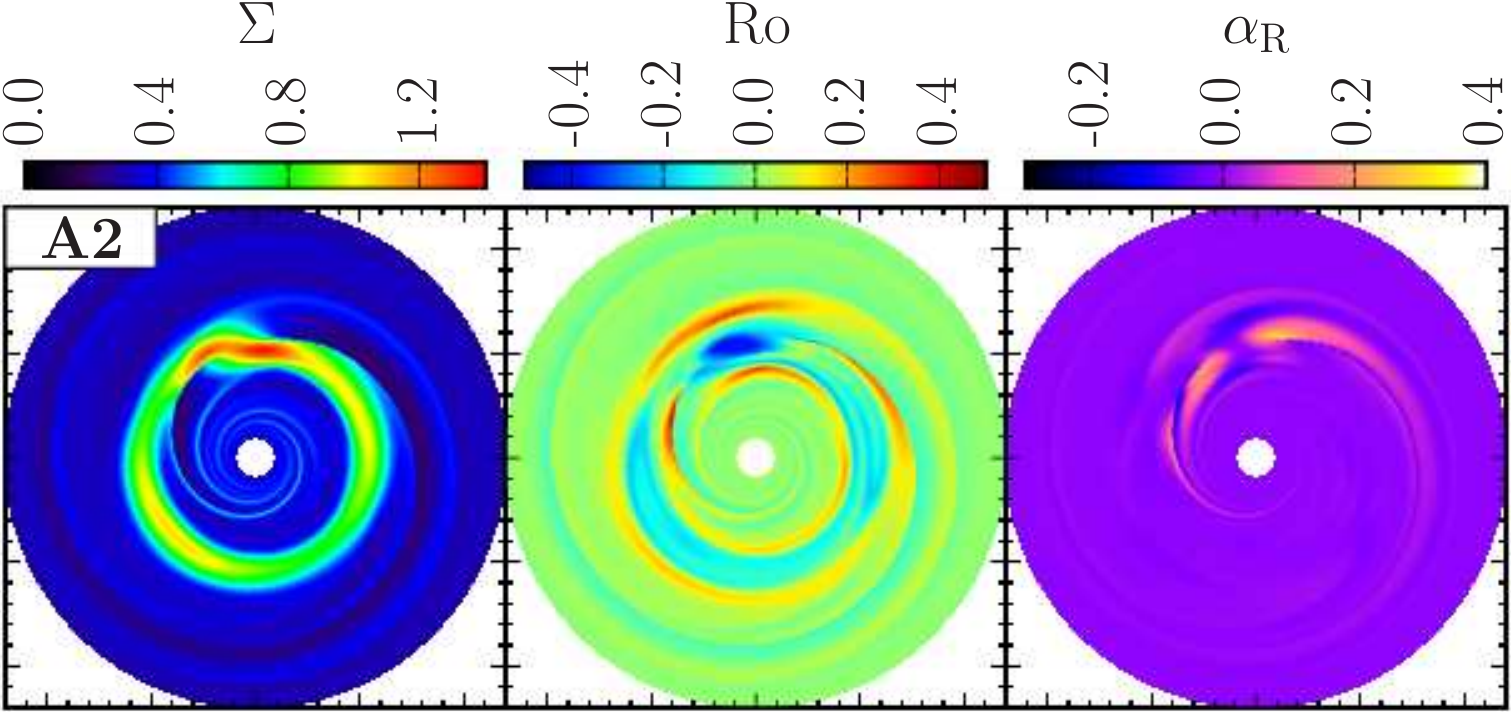}
\caption{Snapshots of surface density, Rossby number, and Reynolds stress, in the quasi-steady state (as in Fig. \ref{fig:snapshot_canonical_steady}), for all runs except the canonical run. The scale of the $x$ and $y$ axes are the same as in Fig. \ref{fig:snapshot_canonical_steady}, but the color scales for the three plotted quantities differ between panels.}
\label{fig:snapshot_runs}
\end{center}
\end{figure*}

\begin{figure*}
\begin{center}
\includegraphics[width=0.99\textwidth,clip]{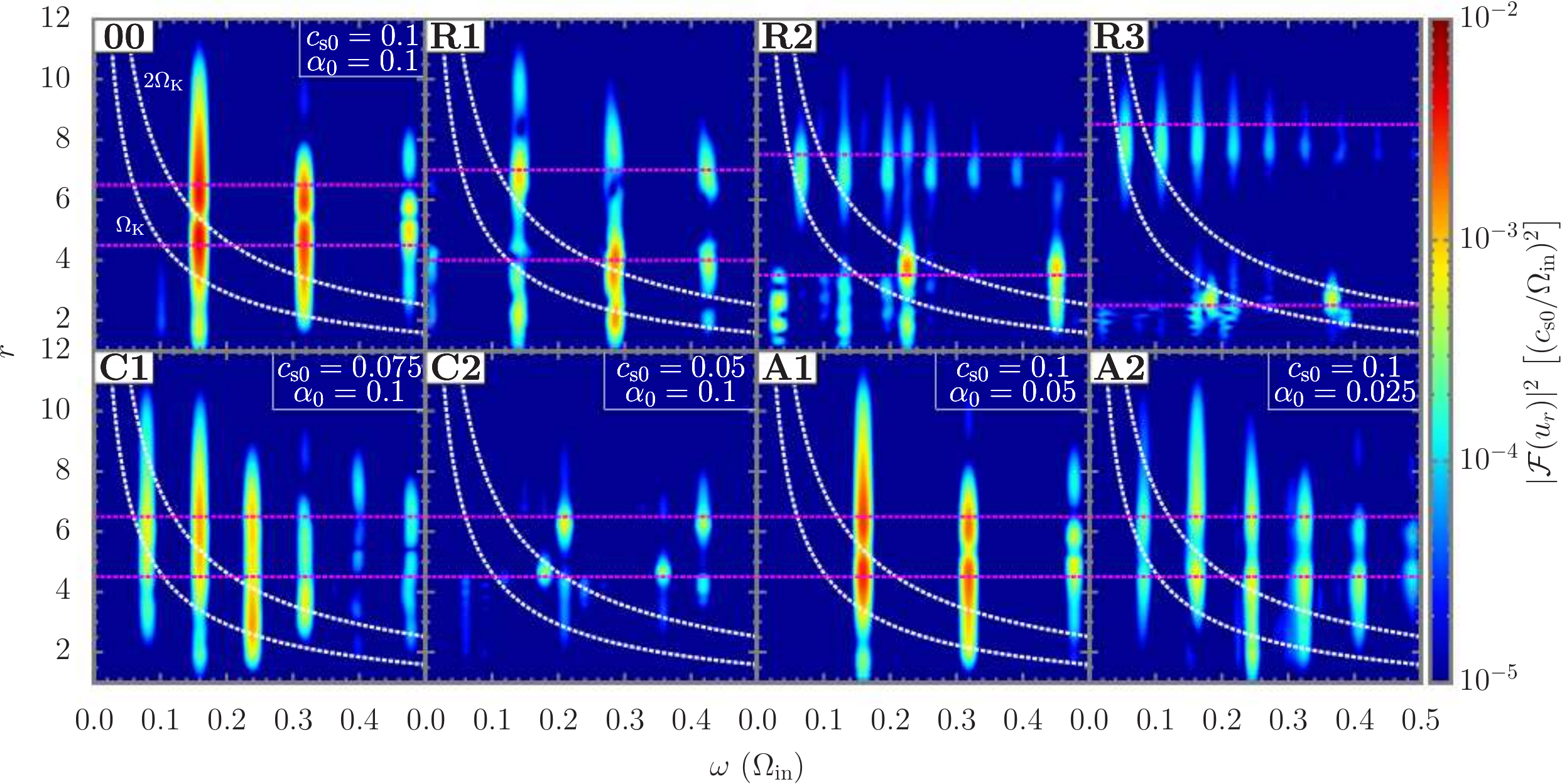}
\caption{Power spectrum of $u_r$ as a function of $r$ for eight different runs. The extent of the DZ, which is varied in the top four panels, is indicated by the horizontal dashed lines. The dashed curves labeled $\Omega_\mathrm{K}$ and $2\Omega_\mathrm{K}$ correspond to the Keplerian orbital frequency and twice the Keplerian orbital frequency.}
\label{fig:psd}
\end{center}
\end{figure*}

\begin{figure}
\begin{center}
\includegraphics[width=0.49\textwidth,clip]{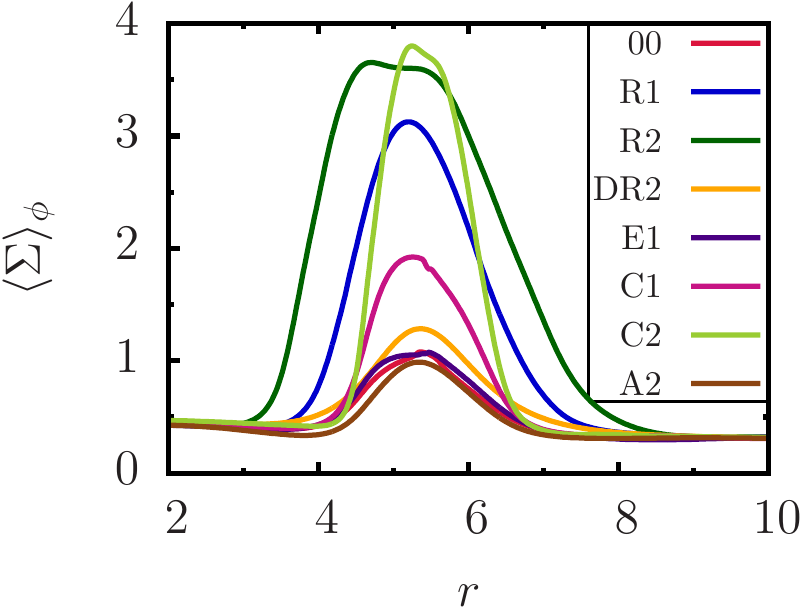}
\caption{Azimuthally-averaged surface density profiles in the quasi-steady state for eight of the eleven runs.}
\label{fig:density_profiles}
\end{center}
\end{figure}

\begin{figure}
\begin{center}
\includegraphics[width=0.49\textwidth,clip]{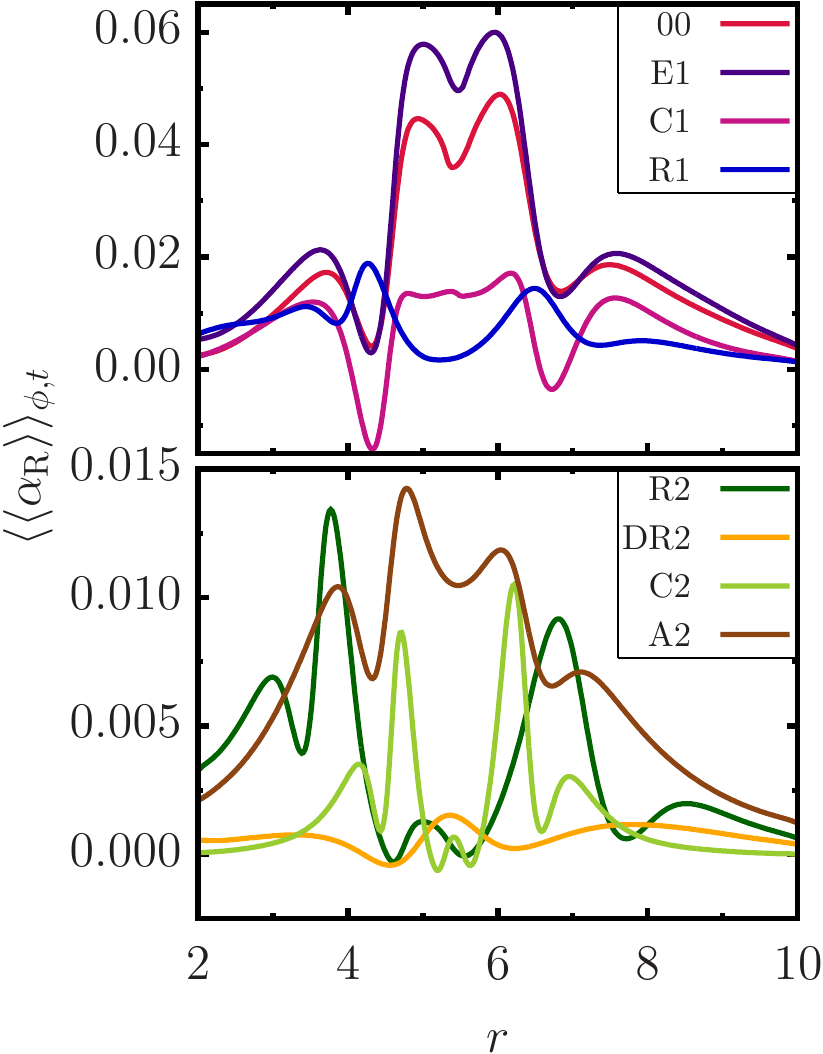}
\caption{Double-averaged Reynolds stress (azimuthally-averaged and time-averaged for $100\,P_\mathrm{in}$) for eight of the eleven runs. Note that the range of the $y$-axis is different in the top and bottom panels. The peak and DZ-averaged values are listed in Table \ref{tab:summary}.}
\label{fig:reynolds_profiles}
\end{center}
\end{figure}

To test the dependence of the angular momentum transport and morphology on the DZ parameters, we ran a suite of simulations, each differing from the canonical run by the value of a single parameter or pair of related parameters. The parameters for each run are listed in Table \ref{tab:summary}. Each run is labeled with letters, indicating which parameter is being varied (R for DZ width, DR for viscosity transition width, E for viscosity reduction factor, C for sound speed, and A for active zone viscosity), and a number, which indexes multiple runs with different parameter values. Most runs, like the canonical run 00, are evolved for $10^4\,P_\mathrm{in}$, but runs R3, A1 and A2 are evolved for $1.5 \times 10^4\,P_\mathrm{in}$, and C2 for $2.5 \times 10^4\,P_\mathrm{in}$. These durations are long enough to allow the disc to reach a quasi-steady state, in which there are no longer appreciable changes in the Fourier components of $\Sigma$, or its azimuthally-averaged profile. While the former typically happens by about $1000\,P_\mathrm{in}$, shortly after the RWI has saturated, the latter does not happen until much later, requiring viscous timescales to reach.

Throughout this section we refer to several figures which illustrate the main results of our parameter study simulations. Figure \ref{fig:snapshot_runs} shows a snapshot of the surface density, Rossby number and Reynolds stress at the end of each simulation (as in Fig. \ref{fig:snapshot_canonical_steady} for run 00). These illustrate the morphology, surface density contrast, presence or absence of vortices, and level of angular momentum transport in the quasi-steady state. Azimuthally-averaged surface density profiles for eight of the eleven simulations are shown in Fig. \ref{fig:density_profiles}, and double-averaged (i.e., averaged over azimuth and time) dimensionless Reynolds stress $\langle\langle\alpha_\mathrm{R}\rangle\rangle_{\phi,t} = \langle\langle\Sigma\tilde{u}_r\tilde{u}_\phi\rangle\rangle_{\phi,t}/(\langle\langle\Sigma\rangle\rangle_{\phi,t}c_\mathrm{s}^2)$ profiles in Fig. \ref{fig:reynolds_profiles}. Together these indicate the efficiency of RWI-driven angular momentum transport (large Reynolds stress and low DZ surface density indicate efficient transport). Figure \ref{fig:psd} illustrates the power spectrum (temporal Fourier transform squared) of $u_r$ taken over $100$ orbits in the quasi-steady state for eight of the simulations. This demonstrates the variety of oscillations present in the disc, the spatial extent of their coherence, and how they are affected by the width of the DZ and by viscosity (as described in Sections \ref{subsec:dz_size} and \ref{subsec:alpha_0}).

Table \ref{tab:summary} summarizes several key quantities characterizing the quasi-steady state for each simulation. The mass contained in the DZ, $M_\mathrm{DZ}$, is expressed in terms of the mass in the DZ at $t = 0$,
\be
\label{eq:mnodz}
M_\mathrm{nodz} = \int_0^{2\pi} \int_{r_\mathrm{IDZ}}^{r_\mathrm{ODZ}} r^{-\frac{1}{2}} r \mathrm{d}r \mathrm{d}\phi,
\ee
which is equal to the mass between $r_\mathrm{IDZ}$ and $r_\mathrm{ODZ}$ for a steady-state axisymmetric disc with $\epsilon_\mathrm{DZ} = 1$, i.e., for a constant-$\alpha$ disc with no DZ (hence the subscript ``nodz''). The maximum surface density $\Sigma_\mathrm{max}$ is given in terms of the surface density in the middle of the DZ at $t = 0$, $\Sigma_{0,\mathrm{DZ}}$ (equal to $1/\sqrt{5.5} = 0.43$ in all runs). This indicates the largest factor by which the surface density is enhanced relative to the ambient density slightly outside of the DZ. The maximum and DZ-averaged values of the double-averaged Reynolds stress, $\langle\langle\alpha_\mathrm{R}\rangle\rangle_{\phi,t}$, computed from the profiles in Fig. \ref{fig:reynolds_profiles}, are also listed in Table \ref{tab:summary}.

\subsection{Dead Zone Width}
\label{subsec:dz_size}
The radial width of the DZ is varied in runs R1, R2 and R3, by decreasing $r_\mathrm{IDZ}$ and increasing $r_\mathrm{ODZ}$. The width of the viscosity transitions, $\Delta r_\mathrm{IDZ/ODZ}$, are scaled with the local scale height as in the canonical run ($\Delta r_\mathrm{IDZ/ODZ} = H/2$), and thus differ in absolute size between these runs. It is useful to define the dimensionless DZ width
\be
\label{eq:delta_dz}
\Delta_\mathrm{DZ} = \frac{2\left(r_\mathrm{ODZ}-r_\mathrm{IDZ}\right)}{r_\mathrm{IDZ}+r_\mathrm{ODZ}},
\ee
i.e., the width of the DZ divided by the the radial coordinate of its center. The canonical run (and all others besides the ``R'' runs) has $\Delta_\mathrm{DZ} = 0.36$, while runs R1, R2 and R3 have $\Delta_\mathrm{DZ} = 0.55, 0.73$ and $1.09$. These four runs demonstrate the features of ``narrow'' and ``wide'' DZs, as well as the intermediate behavior between these two regimes.

The main effect of varying the width of the DZ is shown in the top four panels of Fig. \ref{fig:psd}. Here the power spectrum of radial velocity is shown for runs 00 and R1-R3, with increasing $\Delta_\mathrm{DZ}$ from left to right. For the smallest DZ width (as detailed in Section \ref{sec:canonical_run}), there is a global, coherent oscillation mode with a frequency equal to twice (i.e., an $m = 2$ mode) the Keplerian orbital frequency (evaluated at the DZ center) spanning the entire DZ, as well as its higher frequency harmonics. This represents the typical behavior of a ``narrow'' DZ. For the largest DZ width (run R3), there is a distinct mode present near each DZ edge, each corresponding to the Keplerian frequency at a radius close to the edge (i.e., they are $m = 1$ modes), along with harmonics. Each mode is coherent only near its respective DZ edge, and does not effectively interact with the other edge. This represents a typical ``wide'' DZ. Intermediate separations (runs R1 and R2; the top middle panels of Fig. \ref{fig:psd}) lead to more complicated behaviors in between those of narrow and wide DZs. In these cases, there are two distinct modes, each with a fundamental frequency associated with one DZ edge, but not necessarily with the same $m$. For example, R2 has an $m = 2$ mode near the inner DZ edge [i.e., its frequency is approximately $2\Omega_\mathrm{K}(r_\mathrm{IDZ})$] and an $m = 1$ mode near the outer edge. There are varying degrees of coherence between the edge modes, but they are distinctly less coherent than for the narrow DZ case. The transition between the narrow and wide DZ regimes occurs at $\Delta_\mathrm{DZ} \approx 0.7$.

As the DZ width increases, the regions of negative vorticity become highly azimuthally elongated, and are not coincident with distinct overdensities (see Fig. \ref{fig:snapshot_runs} for runs R1-R3 and Fig. \ref{fig:snapshot_canonical_steady} for run 00). The effective Reynolds stress also decreases (see Table \ref{tab:summary}), with only run R1 having $\langle\alpha_\mathrm{R}\rangle_\mathrm{DZ} \sim 10^{-2}$ (recall that $10^{-2}$ is the residual intrinsic DZ viscosity). Consequently, the mass in the DZ for runs R1-R3 is two to three times larger than in the canonical run.

\subsection{Viscosity Transition Width}
\label{subsec:dz_edgewidth}
We explore the effects of changing the viscosity transition widths in runs DR1 and DR2, in which $\Delta r_\mathrm{IDZ}$ and $\Delta r_\mathrm{ODZ}$ are two and three times larger than in run 00. In its quasi-steady state, DR1 is very similar to the canonical run, in terms of morphology, DZ mass (slightly larger than the canonical run), and peak/DZ-averaged Reynolds stress (slightly smaller than the canonical run). Run DR2 is dramatically different, with the Reynolds stress reduced by two orders of magnitude, and a morphology which resembles that of the linear-phase RWI (e.g., the third panel of Fig. \ref{fig:snapshots_canonical_evolution}), rather than having the strong nonlinear features of the canonical run. The evolution of the $\Sigma_m$'s reveals that their exponential growth is halted at a smaller amplitude than in the canonical run, and subsequently maintained at that amplitude, suggesting that the RWI is partially suppressed by viscosity.

This behavior can be understood as follows. The broader viscosity transitions reduce the sharpness of the density bump which develops in the DZ. This has two consequences on the linear RWI: the intrinsic growth rate is reduced, and the viscous damping rate is also reduced (i.e., the viscous diffusion time across the bump increases). As shown in Appendix \ref{sec:visc_rwi}, competition between these two effects results in the existence of a critical $\alpha$ above which the RWI is suppressed. We estimate that for a bump of width of $H$, $\alpha_\mathrm{crit} \approx 0.04$. Thus, in run DR1, the RWI is not strongly affected by viscosity, since $\alpha_\mathrm{DZ} = 0.01 < \alpha_\mathrm{crit}$. An estimate of $\alpha_\mathrm{crit}$ relevant to run DR2 suggests that the growth of the RWI is significantly reduced by viscosity. Strictly, our viscous damping criterion only applies to the linear RWI, but we expect the nonlinear evolution to be affected to a similar degree. This explains the qualitatively different outcome of run DR2 compared to DR1. We conclude that efficient revival of the DZ by the RWI requires $\Delta r_\mathrm{IDZ}, \Delta r_\mathrm{ODZ} \lesssim H$ (in agreement with previous studies, e.g., Lyra et al. 2009b; Reg\'{a}ly et al. 2012).

\subsection{Viscosity Reduction Factor}
\label{subsec:dz_epsilon}
In run E1, we set $\epsilon_\mathrm{DZ} = 0.01$, ten times smaller than in the canonical run. In an axisymmetric, purely viscously evolving disc, this would result in a surface density enhancement of $\sim \epsilon_\mathrm{DZ}^{-1} = 100$ (compared to a constant $\alpha$ disc) in the DZ. However, once a quasi-steady state has been reached, the mass accumulated in the DZ is only about $5\%$ larger than in the canonical run. All other features of this run are also very similar, such as the azimuthal symmetry and vortex shape (see Fig. \ref{fig:snapshot_runs}). The main difference is the strength of angular momentum transport; the peak and DZ-averaged Reynolds stresses are $\sim 25\%$ larger. This may be due to the fact that the reduced DZ viscosity leads to less viscous damping of the vortex and density wave motions, allowing them to be more vigorous. This behavior is in qualitative agreement with our analysis of the effect of viscosity in the linear regime (see Appendix \ref{sec:visc_rwi}). Overall, compared to other DZ parameters, the value of $\epsilon_\mathrm{DZ}$ has a relatively minor effect on the properties of the quasi-steady state of the disc.

\subsection{Sound Speed}
\label{subsec:dz_cs}
The effects of lowering the sound speed, by means of lowering the aspect ratio $h$, are illustrated in Fig. \ref{fig:psd}. For $h = 0.075$ (run C1), the results are similar to run 00, where a coherent global oscillation develops in the DZ. However, it is dominated by an $m = 1$ mode (i.e., $\omega \approx \Omega_\mathrm{K}$) rather than an $m = 2$ mode. This may be related to the role of viscosity, since $\nu \propto \alpha h^2$, and a similar effect is seen when $\alpha_0$ is reduced (see Section \ref{subsec:alpha_0}). The Reynolds stress is reduced by a factor of three and the mass in the DZ increases by about $75\%$.

When $h$ is further reduced to $0.05$ (run C2), the behavior becomes similar to the wide DZ case described in Section \ref{subsec:dz_size}. Oscillations localized at each DZ edge are present, with azimuthal numbers greater than unity, and with relatively low amplitude in the power spectrum. The Reynolds stress becomes small, about an order of magnitude smaller than in run 00, and the mass in the DZ becomes large (three times larger than run 00) to compensate for the low level of stress. In this case, the wide DZ behavior occurs because the sound speed is too small to allow the entire DZ to oscillate coherently. In order for both DZ edges to interact with one another, the sound crossing time of the DZ, $t_\mathrm{cross} \sim (r_\mathrm{ODZ}-r_\mathrm{IDZ})/\langle c_\mathrm{s} \rangle_\mathrm{DZ}$, should be less than the period of a density wave produced by the RWI, which is approximately the local orbital period, i.e., the ratio $t_\mathrm{cross}/P_\mathrm{orb,DZ} \sim \Delta_\mathrm{DZ}/(2\pi h)$ should be less unity. This criterion is satisfied in run C1, but not in run C2.

\subsection{Active Zone Viscosity}
\label{subsec:alpha_0}
Run A1, in which $\alpha_0$ is reduced to half of its canonical run value, produces a quasi-steady state similar to that of run 00. The morphology is dominated by the same $m = 2$ mode and the shape of the vortex is the same. The DZ mass and maximum surface density are about $20\%$ smaller, and the Reynolds stress (either the peak or DZ-averaged value) is $35\%$ smaller. This suggests that the disc self-regulates the Reynolds stress produced in the DZ, in order to achieve a similar level of angular momentum transport as in the active zone.

Further reducing $\alpha_0$ in run A2 leads to many similar properties in the quasi-steady state. There is a small increase in $M_\mathrm{DZ}$ and $\Sigma_\mathrm{max}$, and a corresponding reduction in $\alpha_\mathrm{R}$. However, there is a major change in the morphology of the disc. The most prominent azimuthal mode number becomes $m = 1$ rather than $m = 2$. A similar morphological transition occurs when the sound speed is reduced (see Section \ref{subsec:dz_cs}). In both cases, this change is associated with increasing the Reynolds number, $\mathrm{Re} \sim r^2\Omega_\mathrm{K}/\nu \sim \alpha_0^{-1}h^{-2}$ (in the active zone; note that the Reynolds number in the DZ does not affect the disc morphology, see Section \ref{subsec:dz_epsilon}), which describes the importance of inertial forces relative to viscous forces. Based on the values of the Reynolds number in runs 00, C1, A1, and A2, we conclude that a narrow DZ ($\Delta_\mathrm{DZ} \lesssim 0.7$) with $\mathrm{Re} \lesssim 1800-2000$ leads to an $m = 2$ morphology, otherwise an $m = 1$ morphology is produced.

The physical origin of the $m = 1$ to $m = 2$ morphological change is not clear. We speculate that it is related to the self-regulation of the stress produced in the DZ in order to match the viscous stress in the active zone. For large viscosities, the stress associated with the $m = 1$ mode is insufficient to match the angular momentum transport in the active zone, and so its amplitude is reduced relative to the $m = 2$ mode, which can sustain stronger transport.

\subsection{Gravitational Stability}

\begin{figure}
\begin{center}
\includegraphics[width=0.49\textwidth,clip]{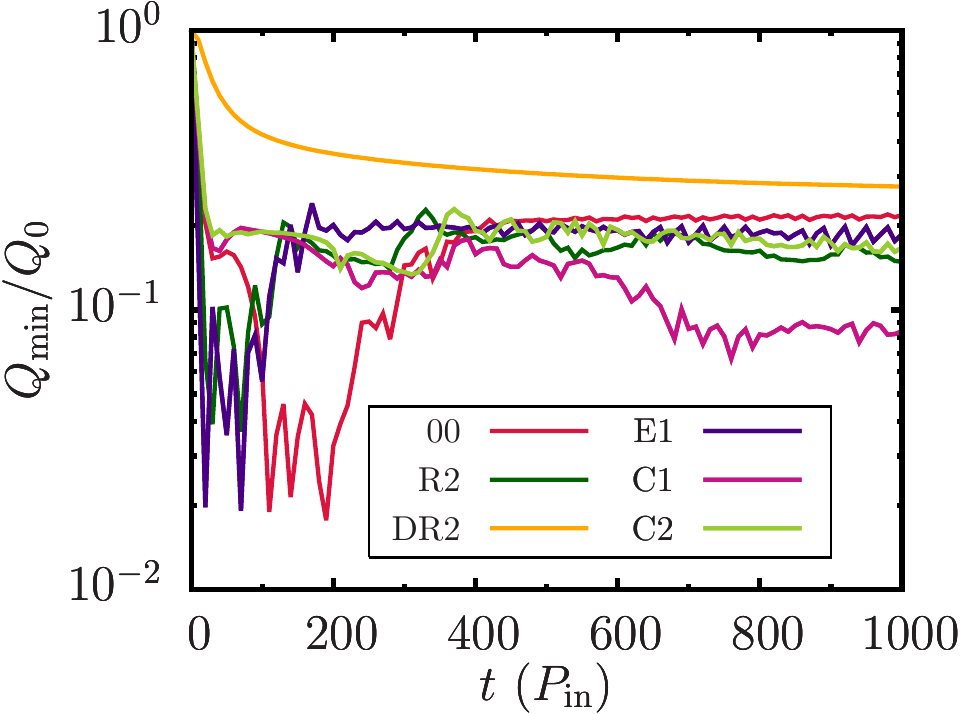}
\caption{Minimum Toomre parameter in the DZ, $Q_\mathrm{min}$, as a function of time, normalized by $Q_0$ [see Eq. (\ref{eq:q_out})], for $t < 1000\,P_\mathrm{in}$ in several runs.}
\label{fig:min_q}
\end{center}
\end{figure}

Self-gravity is not included in our simulations. To evaluate its possible importance, we calculate the local value of Toomre $Q$ parameter,
\be
Q = \frac{\kappa c_\mathrm{s}}{\pi G \Sigma}.
\label{eq:toomre}
\ee
Since in our simulations, $\Sigma$ is scale free, we scale $Q$ with respect to a reference value. For an axisymmetric $\alpha$-disc in steady state, the surface density is related to the accretion rate by $\dot{M} = 3\pi\nu\Sigma$, and the Toomre parameter at radius $r_\mathrm{ODZ}$ is given by
\be
\label{eq:q_out}
\begin{aligned}
Q_0 & = 590 \left(\frac{\alpha}{0.01}\right) \left(\frac{h}{0.1}\right)^3 \left(\frac{r_\mathrm{ODZ}}{10\,\mathrm{AU}}\right)^{-3/2} \\
& \times \left(\frac{M}{1\,M_\odot}\right)^{3/2} \left(\frac{\dot{M}}{10^{-8}\,M_\odot/\mathrm{yr}}\right)^{-1},
\end{aligned}
\ee
which we take as our reference value. We have adopted some fiducial parameters, including a typical accretion rate for protoplanetary discs.

As mass accumulates in the DZ, the $Q$ value in the DZ becomes smaller than $Q_0$. Figure \ref{fig:min_q} shows the minimum local value of $Q$ in the DZ, $Q_\mathrm{min}$, normalized by $Q_0$, as a function of time (for $t < 1000\,P_\mathrm{in}$) for several runs. We see that before the RWI develops, the ratio $Q_\mathrm{min}/Q_0$ can be as small as $0.02$ (e.g., run 00). Once the quasi-steady state is reached, $Q_\mathrm{min}/Q_0$ settles to a modest value between $0.08$ and $0.3$, depending on the parameters of the simulation. The largest reduction ($Q_\mathrm{min}/Q_0 \approx 0.08$) occurs for run C1 (reduced sound speed), which is also the run for which $\Sigma_\mathrm{max}/\Sigma_{0,\mathrm{DZ}}$ is largest (see Table \ref{tab:summary}).

We conclude that, because of the angular momentum transport associated with the nonlinear RWI, the Toomre $Q$ parameter in the DZ can be reduced from the fiducial value $Q_0$ by at most a factor of $12$, for a wide range of DZ parameters. Therefore, Eq. (\ref{eq:q_out}) can be used to estimate the parameters of the steady-state disc (just outside $r_\mathrm{ODZ}$) for which the DZ will remain gravitationally stable, which occurs when $Q_0/12 \gtrsim 1$. For example, for the fiducial values of $h$, $\alpha$, and $r_\mathrm{ODZ}$ in Eq. (\ref{eq:q_out}), stability is guaranteed as long as $\dot{M} \lesssim 5 \times 10^{-7}\,M_\odot\mathrm{yr}^{-1}$. If the outer DZ edge is instead located at $30\,\mathrm{AU}$, $\dot{M} \lesssim 10^{-7}\,M_\odot\mathrm{yr}^{-1}$ is required for stability.

We note that disc self-gravity may affect the RWI even when $Q > 1$. It has been shown that the RWI can be suppressed or modified by self-gravity when $Q \lesssim \pi h^{-1}/2$ (e.g., Lovelace \& Hohlfeld 2013; Zhu \& Baruteau 2015). A self-consistent treatment of self-gravity is required to fully assess its importance relative to RWI in the evolution of DZs.

\section{Discussion}
\label{sec:discussion}

\subsection{Summary of Results}
We have performed long-term, two dimensional hydrodynamic simulations of protoplanetary discs with dead zones (DZs), modeled as regions with reduced $\alpha$-viscosities. We give significant attention to the case of narrow DZs, with radial extent of the order or less than the distance to the central star. We found that the vortices and density waves produced by the Rossby wave instability (RWI), triggered at vortensity gradients naturally arising in a DZ, are capable of partially reviving it. The disc eventually reaches a quasi-steady state, in which angular momentum transport and accretion through the DZ are achieved by a combination of the density bump that induces RWI, and Reynolds stress created by waves and vortices. Because of the latter, the Toomre $Q$ parameter, which determines gravitational stability, is reduced with respect to a constant-$\alpha$ disc by at most a factor of $12$, i.e., $Q \gtrsim Q_0/12$ [see Eq. (\ref{eq:q_out})]. Therefore, RWI can be activated and transport angular momentum through the DZ before it becomes gravitationally unstable, unless the accretion rate is very high. This results in steady accretion, rather than the episodic outburst cycles which may occur when gravito-turbulence transports angular momentum in the DZ (e.g., Zhu et al. 2010a, 2010b; Martin \& Lubow 2011, 2014). Our results presented in this paper suggest that such episodic cycles are possible only for high accretion rates, e.g., $\dot{M} \gtrsim 5 \times 10^{-7}\,M_\odot\mathrm{yr}^{-1}$, for the fiducial disc parameters adopted in Eq. (\ref{eq:q_out}).

We systematically explored the parameters describing the geometry of the DZ and the disc properties, and quantified the transport efficiency of the RWI and the mass enhancement in the DZ (see Table \ref{tab:summary}). In narrow DZs [$\Delta_\mathrm{DZ} \lesssim 0.7$, see Eq. (\ref{eq:delta_dz})], provided that the width of the viscosity transition is not much wider than the local scale height, the azimuthally-averaged Reynolds stresses reach maximum values of $\sim 0.01 - 0.06$, and DZ-averaged Reynolds stresses are in the range $\sim 0.01 - 0.05$. Typically, the mass in the DZ is enhanced relative to a constant-$\alpha$ disc by a factor of two or less, while the density enhancements in the vortices can be as large as large as $4$. For wide DZs, RWI is less efficient, resulting in peak Reynolds stresses $\lesssim 0.02$, and DZ-averaged Reynolds stresses in the range $0.002 - 0.009$. In this case, the mass enhancements in the DZ are about $5 - 7$, with maximum density enhancements reaching $10 - 14$ (relative to constant-$\alpha$ discs).

The morphology of the disc in the quasi-steady state depends on the size of the DZ. For wide DZs, an RWI-unstable vortensity profile is created at either DZ edge, and the resulting vortices merge to produce an $m = 1$ morphology. More interesting phenomena arise for narrow DZs. When the inner and outer DZ edges are sufficiently close, the entire DZ behaves as a single instability site, producing coherent global oscillations. For low viscosities (Reynolds number $\gtrsim 1800 - 2000$), the global wave pattern reaches a nonlinear $m = 1$ mode, consistent with the nonlinear outcome of RWI in the inviscid limit. Increasing the viscosity results in a nonlinear $m = 2$ mode, with two vortices situated at the same radius, separated by $180^\circ$ in azimuth. 

\subsection{Limitations and Prospects}
In our simulations, we modeled the DZ using a radially-dependent viscosity parameter $\alpha(r)$ in a two-dimensional (height-integrated) disc. However, real protoplanetary discs and DZs have three-dimensional, vertically-layered structure (Gammie 1996; Armitage 2011). In a layered disc, MRI-active surface layers of approximately constant surface density $\Sigma_\mathrm{active} \sim 100\,\mathrm{g}/\mathrm{cm}^2$ bookend a dead midplane with surface density $\Sigma_\mathrm{tot} - \Sigma_\mathrm{active}$ that varies with radius. The outer edge of the DZ correponds to the radius beyond which $\Sigma_\mathrm{tot} < \Sigma_\mathrm{active}$, so that the entire column of the disc is MRI-active. Thus, a more realistic description of the DZ involves a viscosity parameter which varies with both radius and height, $\alpha(r,z)$. Since the RWI leads to large fluctuations in surface density, a further refinement would require a parameterization of the form $\alpha(r,z,\Sigma)$, in order to account for fluctuations in the depth of the dead layer. These provide possibilities for future three-dimensional simulations.

We have neglected the role of dust grains, which have several important effects. The accumulation of marginally coupled dust grains (those with drag stopping times similar to the dynamical time) in anticyclonic vortices can aid the formation of planetesimals. Vortices in the outer $\sim 50 - 100$ AU of protoplanetary discs may be responsible for the asymmetries seen in mm-wave observations of transition discs (e.g., van der Marel et al. 2013; Casassus et al. 2013; Isella et al. 2013), since marginally coupled grains at these radii correspond to a size of mm-cm, and thus contribute significantly to mm emission. Vortices may arise due to the presence of DZs, as described in this paper, if there are viscosity transitions at these large radii. Alternatively, they may be produced by gaps opened by massive planets. If viscosity transitions are present in the inner $\sim 10$ AU, structures similar to the ones found in our simulations may be found with future observations which can resolve these scales. The gas morphologies presented in this paper cannot be used as proxies for the dust morphologies, but provide a rough approximation. Simulations of mutually coupled gas and dust (e.g., M\'{e}heut et al. 2012c; Zhu \& Stone 2014) applied to the DZ scenario are required to fully assess this problem.

\section*{Acknowledgments}
This work has been supported in part by NSF grant AST-1211061, and NASA grants NNX14AG94G and NNX14AP31G.

\appendix

\section{Effect of Viscosity on Linear Rossby Wave Instability}
\label{sec:visc_rwi}
Viscosity plays two main roles in our simulations. It drives the evolution of the disc surface density profile, creating RWI-unstable bumps in the DZ, as well as maintaining the bumps as the RWI attempts to smooth them out. It also has a direct effect on the RWI, damping the growth of the linear instability, as well as affecting its nonlinear evolution. Here we investigate the effect of viscosity on the linear growth rate by performing simulations of the RWI on an artificially-imposed density bump.

\begin{figure}
\begin{center}
\includegraphics[width=0.49\textwidth,clip]{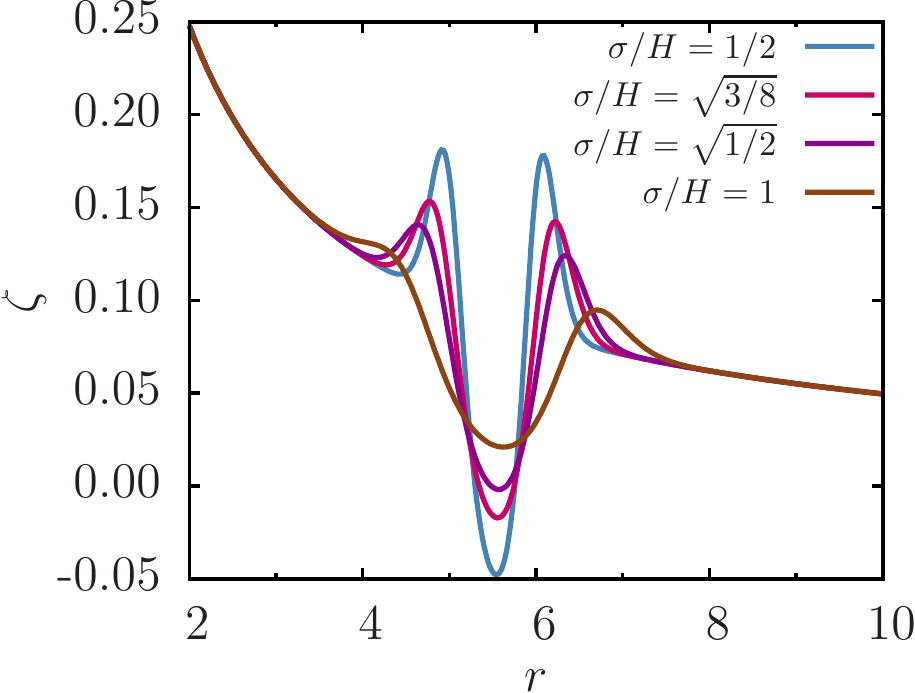}
\caption{Vortensity profiles used for simulations of RWI on a fixed density bump. All profiles have $r_\mathrm{B} = 5.5$ and $\chi = 1$.}
\label{fig:pv_bump}
\end{center}
\end{figure}

We choose a power-law surface density profile modified by a Gaussian bump, given by
\be
\Sigma\left(r\right) = r^{-1/2}\left\{1 + \chi \exp\left[-\frac{\left(r-r_\mathrm{B}\right)^2}{2 \sigma^2}\right]\right\},
\ee
where $r_\mathrm{B}$ is the location of the bump, $\sigma$ is its width and $\chi$ is its dimensionless amplitude. We henceforth set $r_\mathrm{B} = 5.5$ and $\chi = 1$, so that $\sigma$ is the only variable parameter of the bump. The rotation profile $\Omega\left(r\right)$ is modified to satisfy centrifugal balance [Eq. (\ref{eq:centrifugal})] given this profile. The resulting vortensity profiles are shown in Fig. \ref{fig:pv_bump}. In general, viscosity induces a radial drift velocity given by
\be
u_r = \frac{2\Omega}{r^2 \kappa^2 \Sigma} \frac{\mathrm{d}}{\mathrm{d}r} \left(r^3 \nu \Sigma \frac{\mathrm{d}\Omega}{\mathrm{d}r}\right).
\ee
We set $r^3 \nu \Sigma \mathrm{d}\Omega/\mathrm{d} r$ to have a constant value in the disc to ensure $u_r = 0$. This is accomplished by choosing a particular $\nu\left(r\right)$ profile, given by
\be
\label{eq:nu_nodrift}
\nu\left(r\right) = \nu_\mathrm{B} \left(\frac{r}{r_\mathrm{B}}\right)^{-3} \left[\frac{\Sigma\left(r\right)}{\Sigma\left(r_\mathrm{B}\right)}\right]^{-1} \left[\frac{\Omega'\left(r\right)}{\Omega'\left(r_\mathrm{B}\right)}\right]^{-1},
\ee
where $\Omega' = \mathrm{d}\Omega/\mathrm{d}r$, and $\nu_\mathrm{B} = \alpha \left(c_s^2/\Omega_\mathrm{K}\right)_{r_\mathrm{B}}$, so that the viscosity is described by an effective $\alpha$ at $r_\mathrm{B}$. Obviously, Eq. (\ref{eq:nu_nodrift}) is not the standard $\alpha$ prescription for viscosity. This choice of viscosity profile ensures that in a 1D (axisymmetric) simulation, the initial density bump does not diffuse, remaining static on viscous timescales. Therefore in our full 2D simulations, we can test the role of viscosity on the growth of the RWI on a stationary bump. As in our main simulations, the disc has $r_\mathrm{in} = 1$ and $r_\mathrm{out} = 12$. For boundary conditions, we choose all fluid variables to be fixed at their initial values at both boundaries. Damping zones are included interior to $r = 2$ and exterior to $r = 10$, in which all variables are relaxed to their initial values on orbital timescales.

The growth rate $\gamma = \mathrm{Im}\left(\omega\right)$ of an RWI mode (with azimuthal number $m$) can be written as
\be
\gamma = \gamma_0 - \gamma_\nu,
\ee
where $\gamma_0 = \epsilon \Omega$ is the inviscid growth rate (where $\epsilon$ depends on the bump profile and sound speed) and $\gamma_\nu$ is the viscous damping rate. We write
\be
\gamma_\nu = \beta \frac{\nu}{\sigma^2},
\ee
where $\beta$ is of order unity and depends (weakly) on the bump geometry. Near $r_\mathrm{B}$, we have $\nu \approx \alpha c_s^2/\Omega_\mathrm{K} = \alpha H^2 \Omega$, and so
\be
\label{eq:growth_damp}
\gamma = \left[\epsilon - \alpha \beta \left(\frac{\sigma}{H}\right)^{-2}\right] \Omega.
\ee
Therefore, we expect the growth of the RWI to be suppressed when $\alpha > \alpha_\mathrm{crit} = \epsilon \beta^{-1} \left(\sigma/H\right)^2$.

\begin{figure}
\begin{center}
\includegraphics[width=0.49\textwidth,clip]{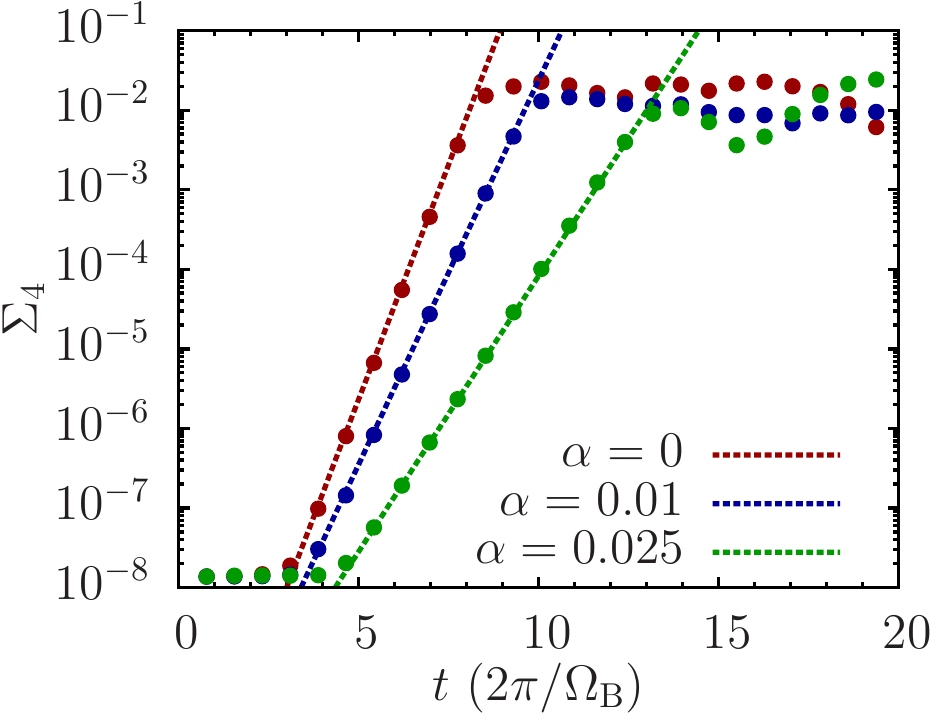}
\caption{Evolution of $\Sigma_4$ [see Eq. (\ref{eq:sigma_m})] for $\sigma = H/2$ and several values of $\alpha$, demonstrating viscous damping of RWI growth. The dashed lines are fits to the linear growth phase, from which the growth rates are determined.}
\label{fig:growthratefits}
\end{center}
\end{figure}

\begin{figure}
\begin{center}
\includegraphics[width=0.49\textwidth,clip]{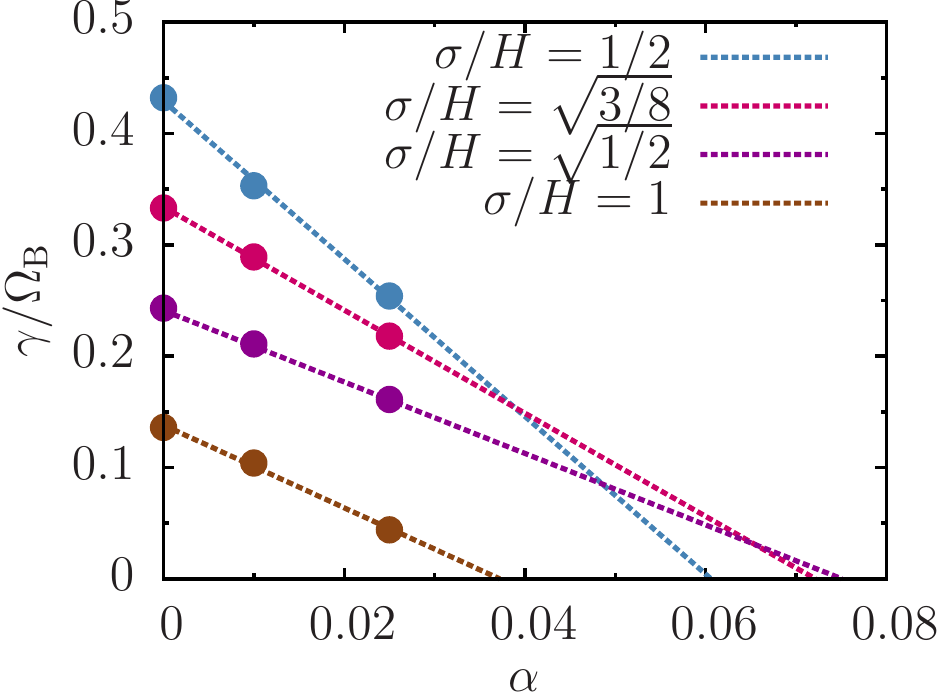}
\caption{Growth rates as as a function of $\alpha$ (points) for different bump widths. The dashed lines are linear fits to the points, whose $x$-intercepts give an estimate of $\alpha_\mathrm{crit}$.}
\label{fig:growthrates}
\end{center}
\end{figure}

We simulated the linear growth of the RWI for four bump widths (see Fig. \ref{fig:pv_bump}), each with $\alpha = 0$, $\alpha = 0.01$, and $\alpha = 0.025$. For each run, we measure the growth rate of the $m = 4$ mode (shown in Fig. \ref{fig:growthratefits} for $\sigma = H/2$). The growth rates as a function of $\alpha$ are shown in Fig. \ref{fig:growthrates}. For each $\sigma$, we find a linear fit to $\gamma(\alpha)$, which determines the value of $\beta$, and hence $\alpha_\mathrm{crit}$, which are given in Table \ref{tab:bump_summary}. For our three narrowest bumps, $\beta$ is nearly constant (approximately $1.6-1.8$), and $\alpha_\mathrm{crit} \approx 0.06 - 0.08$, which is relatively large. For $\sigma = H$, the growth rate is affected more strongly by viscosity ($\beta$ is twice as large), resulting in $\alpha_\mathrm{crit} \approx 0.04$. Therefore, the linear RWI may be significantly affected by viscosity for bumps with $\sigma \gtrsim H$ when $\alpha \sim 10^{-2}$.

\begin{table}
\begin{center}
\begin{tabular}{c|c|c|}
\hline
$\sigma/H$   & $\beta$ & $\alpha_\mathrm{crit}$ \\ \hline
$1/2$        & $1.77$  & $0.0606$               \\
$\sqrt{3/8}$ & $1.74$  & $0.0721$               \\
$\sqrt{1/2}$ & $1.61$  & $0.0751$               \\
$1$          & $3.71$  & $0.0372$               \\ \hline
\end{tabular}
\end{center}
\caption{Viscous damping proportionality factor $\beta$, and critical viscosity parameter $\alpha_\mathrm{crit}$, for different bump widths [see Eq. (\ref{eq:growth_damp})].}
\label{tab:bump_summary}
\end{table}


\begin{thebibliography}{}
\bibitem[]{} Armitage, P. J., 2011, ARA\&A, 49, 195 
\bibitem[]{} Bae, J., Hartmann, L., Zhu, Z., Nelson, R. P., 2014, ApJ, 795, 61 
\bibitem[]{} Bai, X.-N., 2013, ApJ, 772, 96 
\bibitem[]{} Bai, X.-N., 2014a, ApJ, 791, 137 
\bibitem[]{} Bai, X.-N., 2014b, ApJ, 798, 84 
\bibitem[]{} Bai, X.-N., Stone, J. M., 2013, ApJ, 769, 76 
\bibitem[]{} Balbus, S. A., Hawley, J. F., 1991, ApJ, 376, 214
\bibitem[]{} Balbus, S.~A., Hawley, J.~F., 1998, Rev. Mod. Phys., 70, 1 
\bibitem[]{} Balbus, S. A., Papaloizou, J. C. B., 1999, ApJ, 521, 650
\bibitem[]{} Barge, P., Sommeria, J., 1995, A\&A, 295, L1
\bibitem[]{} Bitsch, B., Morbidelli, A., Lega, E., Kretke, K., Crida, A., 2014, A\&A, 570, A75 
\bibitem[]{} Casassus, S., van der Plas, G., M, S. P., Dent, W. R. F., Fomalont, E., Hagelberg, J., Hales, A., Jord\'{a}n, A., Mawet, D., M\'{e}nard, F., Wootten, A., Wilner, D., Hughes, A. M., Schreiber, M. R., Girard, J. H., Ercolano, B., Canovas, H., Rom\'{a}n, P. E., Salinas, V., 2013, Nature, 493, 191
\bibitem[]{} Chang, P., Oishi, J.~S., 2010, ApJ, 721, 1593 
\bibitem[]{} Chiang, E., Youdin, A. N., 2010, AREPS, 38, 493 
\bibitem[]{} Cleeves, L. I., Adams, F. C., Bergin, E. A., 2013, ApJ, 772, 5
\bibitem[]{} Desch, S.~J., Turner, N.~J., 2015, ApJ, 811, 156 
\bibitem[]{} Espaillat, C., Muzerolle, J., Najita, J., Andrews, S., Zhu, Z., Calvet, N., Kraus, S., Hashimoto, J., Kraus, A., D'Alessio, P., 2014, Protostars and Planets VI, University of Arizona Press (arXiv:1402.7103)
\bibitem[]{} Fu, W., Li, H., Lubow, S., Li, S., 2014a, ApJ, 788, L41 
\bibitem[]{} Fu, W., Li, H., Lubow, S., Li, S., Liang, E., 2014b, ApJL, 795, L39 
\bibitem[]{} Gammie, C. F., 1996, ApJ, 457, 355
\bibitem[]{} Gholipour, M., Nejad-Asghar, M., 2014, MNRAS, 441, 1910 
\bibitem[]{} Godon, P., Livio, M., 1999, ApJ, 523, 350 
\bibitem[]{} Godon, P., Livio, M., 2000, ApJ, 537, 396 
\bibitem[]{} Haisch, K.~E., Jr., Lada, E.~A., Lada, C.~J., 2001, ApJL, 553, L153 
\bibitem[]{} Hartmann, L., Calvet, N., Gullbring, E., D'Alessio, P, 1998, ApJ, 495, 385 
\bibitem[]{} Isella, A., P\'{e}rez, L. M., Carpenter, J. M., Ricci, L., Andrews, S., Rosenfeld, K., 2013, ApJ, 775, 30
\bibitem[]{} Klahr, H.~H., Bodenheimer, P., 2003, ApJ, 582, 869 
\bibitem[]{} Kretke, K.~A., Lin, D.~N.~C., 2007, ApJL, 664, L55
\bibitem[]{} Lesur, G., Kunz, M. W., Fromang, S., 2014, A\&A, 566, A56 
\bibitem[]{} Li, H., Colgate, S.A., Wendroff, B., Liska, R., 2001, ApJ, 551, 874 
\bibitem[]{} Li, H., Finn, J. M., Lovelace, R. V. E., Colgate, S. A., 2000, ApJ, 533, 1023 
\bibitem[]{} Lin, M.-K., 2014, MNRAS, 437, 575 
\bibitem[]{} Lodato, G., Rice, W.~K.~M., 2004, MNRAS, 351, 630 
\bibitem[]{} Lovelace, R. V. E., Li, H., Colgate, S. A., Nelson, A. F., 1999, ApJ, 513, 805 
\bibitem[]{} Lovelace, R.~V.~E., Hohlfeld, R.~G., 2013, MNRAS, 429, 529 
\bibitem[]{} Lovelace, R. V. E., Romanova, M. M., 2014, Fluid Dyn. Res., 46, 041401 
\bibitem[]{} Lyra, W., Johansen, A., Zsom, A., Klahr, H., Piskunov, N., 2009, A\&A, 497, 869 
\bibitem[]{} Lyra, W., Lin, M.-K., 2013, ApJ, 775,
\bibitem[]{} Lyra, W., Mac Low, M.-M., 2012, ApJ, 756, 62
\bibitem[]{} Lyra, W., Turner, N., McNally, C., 2015, A\&A, 574, A10 
\bibitem[]{} van der Marel, N., van Dishoeck, E. F., Bruderer, S., Birnstiel, T., Pinilla, P., Dullemond, C. P., van Kempen, T. A., Schmalzl, M., Brown, J. M., Herczeg, G. J., Mathews, G. S., Geers, V., 2013, Science, 340, 1199
\bibitem[]{} Martin, R. G., Lubow, S. H., 2011, ApJ, 740, L6
\bibitem[]{} Martin, R. G., Lubow, S. H., 2014, MNRAS, 437, 682
\bibitem[]{} M\'{e}heut, H., Casse, F., Varni\`{e}re, P., Tagger, M., 2010, A\&A 516, 31 
\bibitem[]{} M\'{e}heut, H, Yu, C., Lai, D., 2012a, MNRAS, 422, 2399 
\bibitem[]{} M\'{e}heut, H., Keppens, R., Casse, F., Benz, W., 2012b, A\&A, 542, 9 
\bibitem[]{} M\'{e}heut, H., Meliani, Z., Varniere, P., Benz, W., 2012c, A\&A, 545, A134 
\bibitem[]{} M\'{e}heut, H., Lovelace, R.~V.~E., Lai, D., 2013, MNRAS, 430, 1988
\bibitem[]{} Mignone, A., Bodo, G., Massaglia, S., Matsakos, T., Tesileanu, O., Zanni, C., Ferrari, A., 2007, ApJS, 170, 228
\bibitem[]{} Rafikov, R.~R., 2015, ApJ, 804, 62 
\bibitem[]{} Reg\'{a}ly, Z., Juh\'{a}sz, A., S\'{a}ndor, Z., Dullemond, C. P., 2012, MNRAS, 419, 1701 
\bibitem[]{} Russo, M., Thompson, C., 2015, ApJ, 813, 81 
\bibitem[]{} Sano, T., Miyama, S. M., Umebayashi, T., Nakano, T., 2000, ApJ, 543, 486 
\bibitem[]{} Tanga, P., Babiano, A., Dubrulle, B., Provenzale, A., 1996, Icarus, 121, 158
\bibitem[]{} Turner, N. J., Fromang, S., Gammie, C., Klahr, H., Lesur, G., Wardle, M., Bai, X.-N., 2014, Protostars and Planets VI, University of Arizona Press (arXiv:1401.7306) 
\bibitem[]{} Urpin, V., Brandenburg, A., 1998, MNRAS, 294, 399 
\bibitem[]{} Varni\`{e}re, P., Tagger, M., 2006, A\&A, 446, 13
\bibitem[]{} Zhu, Z., Hartmann, L., Gammie, C. F., 2010a, ApJ, 713, 1143
\bibitem[]{} Zhu, Z., Hartmann, L., Gammie, C. F., Book, L.G., Simon, J.B., Engelhard, E., 2010b, ApJ, 713, 1134
\bibitem[]{} Zhu, Z., Stone, J. M., 2014, ApJ, 795, 53
\bibitem[]{} Zhu, Z., Baruteau, C., 2015, arXiv:1511.03497 
\end{thebibliography}
\end{document}